%% file: dd89arxiv.tex
\documentclass[12pt]{article}
\usepackage{longtable}
\usepackage[section]{placeins}
\usepackage[top=1in, left=1in, right=1in, bottom=1in]{geometry}
\usepackage[parfill]{parskip}	% Activate to begin paragraphs with an empty line rather than an indent
\usepackage{graphicx}
\usepackage{caption}
\usepackage{subcaption}
\usepackage[export]{adjustbox}
\usepackage{amssymb}
\usepackage{epstopdf}
\usepackage{bbm}
\usepackage{bm}
\usepackage{amssymb}
\usepackage{amsmath}
\usepackage{amsfonts}

\usepackage{lscape}
\usepackage{rotating}
\usepackage{setspace}
\usepackage{threeparttable}
\usepackage{booktabs}
\usepackage[compact]{titlesec}
\usepackage{lmodern}
\fontfamily{lmtt}\selectfont
\usepackage[T1]{fontenc}
\newcommand{\argmax}{\arg\!\max}
\usepackage[nocitation]{apacite}
\usepackage{natbib}
\bibpunct{(}{)}{;}{a}{}{,}
\usepackage{hyperref}
\usepackage{titling}
\usepackage{pifont}
\usepackage[capposition=top]{floatrow}
\newcommand{\subtitle}[1]{%
  \posttitle{%
    \par\end{center}
    \begin{center}\large#1\end{center}
    \vskip0.5em}%
}
\usepackage{amsthm}
\usepackage{amsmath} %maths
\usepackage{multirow}
\usepackage{mathrsfs}

\usepackage{amssymb}% http://ctan.org/pkg/amssymb
\usepackage{pifont}% http://ctan.org/pkg/pifont
\newcommand{\xmark}{\ding{55}}%

\usepackage{changepage}
\newenvironment{widerequation}{%
    \begin{adjustwidth}{-.45cm}{-.45cm}\begin{equation}}
    {\end{equation}\end{adjustwidth}}
    
\newcommand\independent{\protect\mathpalette{\protect\independenT}{\perp}}
\def\independenT#1#2{\mathrel{\rlap{$#1#2$}\mkern2mu{#1#2}}}

\newtheorem*{remark*}{Remark}

\usepackage{xpatch}
\makeatletter
\xpatchcmd{\@thm}{\thm@headpunct{.}}{\thm@headpunct{}}{}{}
\makeatother

\usepackage{mathrsfs}

\usepackage{color}
\begin{document}
\pagestyle{plain}

\newtheoremstyle{mystyle}% name
{\topsep}% Space above
{\topsep}% Space below
{\it}% Body font
{}% Indent amount
{\bf}% Theorem head font
{.}%Punctuation after theorem head
{.5em}%Space after theorem head
{}% theorem head spec
\theoremstyle{mystyle}
\newtheorem{assumptionex}{Assumption}
\newenvironment{assumption}
  {\pushQED{\qed}\renewcommand{\qedsymbol}{}\assumptionex}
  {\popQED\endassumptionex}
\newtheorem{assumptionexp}{Assumption}
\newenvironment{assumptionp}
  {\pushQED{\qed}\renewcommand{\qedsymbol}{}\assumptionexp}
  {\popQED\endassumptionexp}
\renewcommand{\theassumptionexp}{\arabic{assumptionexp}$'$}

\newtheorem{assumptionexpp}{Assumption}
\newenvironment{assumptionpp}
  {\pushQED{\qed}\renewcommand{\qedsymbol}{}\assumptionexpp}
  {\popQED\endassumptionexpp}
\renewcommand{\theassumptionexpp}{\arabic{assumptionexpp}$''$}

\newtheorem{assumptionexppp}{Assumption}
\newenvironment{assumptionppp}
  {\pushQED{\qed}\renewcommand{\qedsymbol}{}\assumptionexppp}
  {\popQED\endassumptionexppp}
\renewcommand{\theassumptionexppp}{\arabic{assumptionexppp}$'''$}

\renewcommand{\arraystretch}{1.3}

\newcommand{\argmin}{\mathop{\mathrm{argmin}}}
\makeatletter
\newcommand{\grande}{\bBigg@{2.25}}
\newcommand{\enorme}{\bBigg@{5}}

\newcommand{\blind}{0}

\newcommand{\tit}{\Large Complex Discontinuity Designs Using Covariates: Impact of School Grade Retention on Later Life Outcomes in Chile}

\if0\blind

{\title{\tit\thanks{The authors thank Zach Branson, Kosuke Imai, Luke Keele, Yige Li, Luke Miratrix, Bijan Niknam, Paul Rosenbaum, Zirui Song, Stefan Wager, the Editor, and two anonymous reviewers for helpful comments and inputs.
The authors were supported in part by award ME-2019C1-16172 from the Patient-Centered Outcomes Research Institute
(PCORI) and grant G-2018-10118 from the Alfred P. Sloan Foundation.}}
\author{Juan D. D\'{i}az\thanks{Department of Management Control and Information Systems,
 Faculty of Economics and Business, University of Chile, Diagonal Paraguay 257, Piso 21, Santiago, Chile (c\'odigo postal, 8330015); email: \url{juadiaz@fen.uchile.cl}.} \and Jos\'{e} R. Zubizarreta\thanks{Departments of Health Care Policy, Biostatistics, and Statistics, Harvard University, 180 A Longwood Avenue, Office 307-D, Boston, MA 02115; email: \url{zubizarreta@hcp.med.harvard.edu}.}
}

\date{}

\maketitle
}\fi

\if1\blind
\title{\bf \tit}
\date{}
\maketitle
\fi

\begin{abstract}
\input{dd89_sec0_01}
\end{abstract}

\begin{center}
\end{center}
\clearpage
\doublespacing

%\singlespacing
%\pagebreak
%\tableofcontents
%\pagebreak
%\doublespacing

%%%%%%%%%%%%%%%%%%%%%%%%%%%%%%%%%%%%%%%%%%%%%
%%%%%%%%%%%%%%%%%%%%%%%%%%%%%%%%%%%%%%%%%%%%%
%%%%%%%%%%%%%%%%%%%%%%%%%%%%%%%%%%%%%%%%%%%%%
\section{Introduction}
\label{sec_introduction}
\input{dd89_sec1_01}
\section{Educational and criminal administrative records}
\label{sec_educational}
\input{dd89_sec2_01}

%%%%%%%%%%%%%%%%%%%%%%%%%%%%%%%%%%%%%%%%%%%%
%%%%%%%%%%%%%%%%%%%%%%%%%%%%%%%%%%%%%%%%%%%%
%%%%%%%%%%%%%%%%%%%%%%%%%%%%%%%%%%%%%%%%%%%%
\section{A framework for complex discontinuity designs}
\label{sec_framework}
\input{dd89_sec3_01}

%%%%%%%%%%%%%%%%%%%%%%%%%%%%%%%%%%%%%%%%%%%%
%%%%%%%%%%%%%%%%%%%%%%%%%%%%%%%%%%%%%%%%%%%%
%%%%%%%%%%%%%%%%%%%%%%%%%%%%%%%%%%%%%%%%%%%%
\section{Estimation and inference}
\label{sec_estimation}
\input{dd89_sec4_01}

%%%%%%%%%%%%%%%%%%%%%%%%%%%%%%%%%%%%%%%%%%%%
%%%%%%%%%%%%%%%%%%%%%%%%%%%%%%%%%%%%%%%%%%%%
%%%%%%%%%%%%%%%%%%%%%%%%%%%%%%%%%%%%%%%%%%%%
\section{Selection of the neighborhood}
\label{sec_selection}
\input{dd89_sec5_01}
\section{Impact of grade retention on later life outcomes}
\label{sec_results}
\input{dd89_sec6_01}

%%%%%%%%%%%%%%%%%%%%%%%%%%%%%%%%%%%%%%%%%%%%
%%%%%%%%%%%%%%%%%%%%%%%%%%%%%%%%%%%%%%%%%%%%
%%%%%%%%%%%%%%%%%%%%%%%%%%%%%%%%%%%%%%%%%%%%
\section{Further extensions}
\label{sec_further_extensions}

\input{dd89_sec7_01}
\section{Concluding remarks}
\label{sec_concluding}
\input{dd89_sec8_01}

%%%%%%%%%%%%%%%%%%%%%%%%%%%%%%%%%%%%%%%%%%%
%%%%%%%%%%%%%%%%%%%%%%%%%%%%%%%%%%%%%%%%%%%
%%%%%%%%%%%%%%%%%%%%%%%%%%%%%%%%%%%%%%%%%%%
%\pagebreak
\onehalfspacing
\bibliography{mybibliography18}
\bibliographystyle{asa}

%%%%%%%%%%%%%%%%%%%%%%%%%%%%%%%%%%%%%%%%%%%%
%%%%%%%%%%%%%%%%%%%%%%%%%%%%%%%%%%%%%%%%%%%%
%%%%%%%%%%%%%%%%%%%%%%%%%%%%%%%%%%%%%%%%%%%%
\pagebreak
\setcounter{page}{1}
\section*{Online supplementary materials}
\label{sec_concluding}
\input{dd89arxiv_supplementary.tex}

\end{document}

%% file: dd89_sec0_01.tex
Regression discontinuity designs are extensively used for causal inference in observational studies.
However, they are usually confined to settings with simple treatment rules, determined by a single running variable, with a single cutoff.
%In this paper, we review existing methods for multiple running variables and multiple cutoffs, and propose a framework for complex discontinuity designs that simultaneously encompasses them into multiple treatment rules.
%These rules may be determined by multiple running variables, each with many cutoffs, that possibly lead to the same treatment.
Motivated by the problem of estimating the impact of grade retention on educational and juvenile crime outcomes \textcolor{black}{in Chile,} we propose a framework and methods for complex discontinuity designs that encompasses multiple treatment rules.
%In this paper, we review existing methods for multiple running variables and multiple cutoffs, and propose a framework for complex discontinuity designs that simultaneously encompasses them.
%\textcolor{black}{In these designs, treatment assignment rules may be determined by multiple running variables, each with many cutoffs, that possibly lead to the same treatment.}
%Moreover, the running variables may be discrete.
In this framework, the observed covariates play a central role for identification, estimation, and generalization of causal effects.
Identification is non-parametric and relies on a local strong ignorability assumption.
%; that is, on local unconfoundedness and local positivity assumptions.
Estimation proceeds as in any observational study under strong ignorability, yet in a neighborhood of the cutoffs of the running variables.
We discuss estimation approaches based on matching and weighting, including complementary regression modeling adjustments.
We present assumptions for generalization; that is, for identification and estimation of average treatment effects for target populations.
% beyond the study sample that reside in a neighborhood of the cutoffs.
%We also propose two approaches to select the neighborhood for the analyses and assess the plausibility of the assumptions.
We also describe two approaches to select the neighborhood for analysis.
We find that grade retention \textcolor{black}{in Chile} has a negative impact on future grade retention, but is not associated with dropping out of school or committing a juvenile crime.

%% file: dd89_sec1_01.tex
%%%%%%%%%%%%%%%%%%%%%%%%%%%%%%%%%%%%%%%%%%%
%%%%%%%%%%%%%%%%%%%%%%%%%%%%%%%%%%%%%%%%%%%
\subsection{Impact of grade retention \textcolor{black}{in Chile}}
\label{sec_motivating_example}

What is the impact of childhood grade retention on later life outcomes?
This is a central question in education where two visions compete.
%two theories compete
From one perspective, \textcolor{black}{through a negative impact on motivation and self esteem, which may in turn reduce the effectiveness of educational inputs,} repeating a grade is associated with dropout later in schooling and other negative long-term effects.
Alternatively, children who are lagging behind may benefit from repeating a grade by learning the content that they have missed and fitting better with younger peers.

In Chile, for example, school grades vary between 1 and 7, by increments of 0.1.
In this grade system, 7 stands for ``Outstanding,'' 4 denotes ``Sufficient,'' and 1 is ``Very Deficient.''
The child fails a subject with a grade below 4.0.
They repeat the year under either of the following two rules: (1) having a grade below 4 in one subject \emph{and} having an average grade across all subjects lower than 4.5; \emph{or} (2) having a grade below 4 in two subjects \emph{and} having an average grade across all subjects lower than 5.
%In this context, the specific the question that motivates the review and framework presented in this paper is what is the effect of grade retention during childhood on later educational and criminal outcomes.
In this context, we ask what is the effect of grade retention during childhood on later educational and criminal outcomes.

To answer this question, one option is to compare students who passed to those who repeated by carefully adjusting for differences in educational and socioeconomic background characteristics or covariates.
However, this comparison will likely be biased by differences in covariates between the two groups that we fail to observe such as the students' ability and motivation.
Manifestly, in this question there are notions of proximity and chance, in which two similar students (with similar past grades and socioeconomic backgrounds) have marginally different results in a school subject, yet one ends up barely passing the grade while the other must repeat, in a haphazard way.
These notions evoke the idea of a regression discontinuity design, where treated and control units are compared in a vicinity of a threshold to receive treatment.
However, to our knowledge, existing methods do not readily apply to this and other related problems that simultaneously involve multiple treatment assignment variables and thresholds.

%\textcolor{black}{... that simultaneously involve multiple running variables and multiple cutoffs possibly leading to the same treatment.}
%TRATAR DE EXPLICAR CON EL LENGUAJE MAS GENERAL DE INFERENCIA CAUSAL

%In particular, we measure grade retention during elementary education, when students generally take the same subjects and retention rules (1) and (2) apply nationally.
%In Chile, students are evaluated within schools (for this reason, our outcome comparisons are within schools and school grades, among other covariates, by exact matching; see Section \ref{sec_finding} for details).
%We measure educational and criminal outcomes over a 10-year period after grade retention.

%%%%%%%%%%%%%%%%%%%%%%%%%%%%%%%%%%%%%%%%%%%
%%%%%%%%%%%%%%%%%%%%%%%%%%%%%%%%%%%%%%%%%%%
%\subsection{A review of discontinuity designs \textcolor{black}{with complex treatment assignment rules}}
\subsection{Review of discontinuity designs}
\label{sec_review}

\subsubsection{Elements and frameworks}
The regression discontinuity design \citep{thistlethwaite1960regression} or, simply, the discontinuity design, is widely recognized as one of the strongest designs for causal inference in observational studies.
In a discontinuity design, the treatment assignment is governed by a covariate called the assignment, forcing, or running variable, such that for values of this covariate greater or smaller than a given cutoff, subjects are assigned to treatment or control.
The basic intuition behind this design is that subjects just below the cutoff (who are not assigned to treatment) are good counterfactual comparisons to those just above the cutoff (who are assigned to treatment).
In this design, treatment effects are essentially estimated by contrasting weighted average outcome values across treatment groups at the cutoff or in a small neighborhood around it.
See, for example, the reviews by \cite{imbens2008regression}, \cite{lee2014regression}, and \citeauthor{cattaneo2019practical} (\citeyear{cattaneo2019practical}, \citeyear{cattaneo2020practical}).

Following \citeauthor{cattaneo2019practical} (\citeyear{cattaneo2019practical}, \citeyear{cattaneo2020practical}), there are two frameworks for interpreting and analyzing discontinuity designs: the continuity-based framework, which is asymptotic and identifies the effect of treatment at the cutoff  (e.g., \citealp{hahn2001identification,calonico2014robust,gelman2018high,imbens2019optimized}), and the local randomization framework, which is limitless, and formulates the design as a local randomized experiment around the cutoff (e.g., \citealp{cattaneo2015randomization,li2015evaluating,mattei2016regression,sales2020limitless}).
%%While both frameworks have strengths, they are usually confined to settings with simple treatment rules, determined by a single continuous running variable, with a single cutoff, and where covariates other than the running variable are not simultaneously used to adjust and test for covariate balance in order to select the neighborhood for analysis.
%In both frameworks, most applications involve a single running variable with a single cutoff.
%However, many policies rely on more than one such variables and thresholds to determine treatment assignment.
%Two important cases are addressed under both frameworks: the sharp design, where the treatment assignment rule is deterministic, and the fuzzy design, where the rule is probabilistic (for instance, if there is imperfect compliance to treatment assignment).
%In both frameworks, most applications involve a single running variable with a single cutoff; however, many policies rely on more than one such variables and thresholds to determine treatment assignment.
%%\textcolor{black}{Two important cases are addressed under both frameworks: the sharp design, where there is perfect compliance with the treatment assignment rule, and the fuzzy design, when noncompliance with the rule occurs. 
%%In both frameworks, most applications involve a single running variable with a single cutoff; however, many policies rely on more than one such variables and thresholds to determine treatment assignment.}
Two important cases are addressed under both frameworks: the sharp design, where the treatment assignment rule is deterministic, and the fuzzy design, where the rule is not deterministic (e.g., if there is imperfect compliance with the treatment assignment).
In both frameworks, most applications involve a single running variable with a single cutoff; however, as we describe in Section \ref{sec_other_areas}, many policies actually rely on more than one such variables and thresholds to determine treatment assignment.

\subsubsection{Discontinuity designs with multiple cutoffs}
To the best of our knowledge, there are methods that separately incorporate multiple cutoffs and multiple running variables, but not both simultaneously.
One set of methods encompasses single running variables with multiple cutoffs; the other, multiple running variables each with a single cutoff.
Examples of the former include \cite{chay2005multiple}, who estimate the effect of government school funding on students' achievement in Chile, where a school test score cutoff that varies across geographic regions determines funding eligibility; \cite{egger2010rd}, who analyze the effect of the size of city government councils on municipal expenditures in Germany, where the size of the council depends on population cutoffs at the municipality level; and \cite{delamaza2012rd}, who studies the impact of Medicaid benefits on health care utilization, where household income cutoffs determine Medicaid benefits.

A widespread analytic approach in the presence of multiple cutoffs is to pool the data from the multiple cutoffs to produce a single effect estimate.
%standardize the different cutoffs at zero and estimate one effect.
More specifically, in this approach the running variable of each observation is centered around its corresponding cutoff, the data is pooled across the observations, and a single treatment effect estimate is obtained under the continuity-based framework via local linear regressions (e.g., \citealp{cattaneo2020practical}).
By pooling the information across different cutoffs, this estimator can have improved efficiency over the analogous estimator at each cutoff, yet at the cost of using the same kernel and bandwidth for all the cutoffs (see \citealp{cattaneo2016interpreting} for an analysis and interpretation of this pooled estimator under different assumptions).
An alternative approach leverages the multiple cutoffs to estimate more general average treatment effects than those for the individuals observed around the cutoffs.
For instance, \cite{bertanha2020multiple} proposes an estimator for the average treatment effect when heterogeneity is explained by certain cutoffs characteristics.

\subsubsection{Discontinuity designs with multiple running variables}
In discontinuity designs with multiple running variables and single cutoffs, most existing methods again fall under the continuity-based framework and use local polynomial regression methods.
Several of these developments have been motivated by problems in educational and in geographic settings.
For example, \cite{jacob2004rd} study the effect of grade retention on student achievement, where retention is determined by cutoffs in both math and reading exams; and \cite{matsudaira2008multiple} analyzes the impact of summer school on student attainment, where summer school attendance is defined by cutoffs in two standardized tests.
In geographic settings, this design arises when units are assigned to treatment on the basis of a geographic boundary.
For instance, using latitude and longitude as running variables, \cite{keele2015geographic} analyze the impact of political advertisements on voter turnout, and \cite{branson2019nonparametric} study the effect of school district location on house prices (see also \citealp{keele2017overview}).
Naturally, in these settings the selection of the neighborhood and other relevant parameters such as the kernel function is more difficult than in one-dimensional settings (see, e.g., \citealp{imbens2019optimized} for a discussion).
Some applied works avoid these choices by restricting the data (e.g., \citealp{jacob2004rd, matsudaira2008multiple}) or aggregating the running variables into a single dimension (e.g., \citealp{cattaneo2020practical}) to then apply local polynomial regression methods for a single running variable.
Alternatively, \cite{imbens2019optimized} directly minimize a bound of the conditional mean-squared error of a weighting estimator via explicit optimization.
As the authors discuss, an important aspect of this method is the selection of two tuning parameters, which can be challenging with multiple running variables.

\subsubsection{Discontinuity designs with discrete running variables}
In discontinuity designs, an added complication arises when the running variables are discrete. 
\textcolor{black}{The literature distinguishes between two cases, where the running variable is truly discrete (as in our study), and where this variable is rounded or measured with error (see \citealp{pei2017devil} and \citealp{bartalotti2020correction}).
Examples of discontinuity designs with discrete running variables} include \cite{card2004rd}, who estimate the impact of Medicaid expansions on the health insurance status of low-income children, where a child age cutoff determines Medicaid coverage, and reported children age is rounded in years; \textcolor{black}{\cite{almond2010estimating} and \cite{barreca2011saving}, who study the impact of low birth weight classification on infant mortality, where birth weight is rounded to different gram and ounce multiples;} and \cite{manacorda2012rd} who analyzes the impact of grade retention on students' achievement, where retention is mandatory for students that fail three or more school subjects, and the running variable is the number of failed subjects.
Under the continuity-based framework, \cite{dong2015rd} 
shows that estimation using a discrete running variable leads to inconsistent estimates of treatment effects at the cutoff, even when the true regression function is known and is correctly specified. 
This implies that when the running variable takes only a few distinct values, the methods for estimation, inference, and bandwidth selection under the continuity-based framework via local polynomials do not apply. 
See \cite{kolesar2018rd} who propose methods for constructing confidence intervals with guaranteed coverage properties when the running variable takes a moderate number of distinct values (see also \citealp{lee2008rd}). 
As \cite{cattaneo2020practical} argue, the local randomization framework is more natural for analysis in this case.

%To the best of our knowledge, existing approaches to discontinuity designs do not accommodate such complex treatment assignment rules with
%
%simultaneously multiple discrete running variables taking a few distinct values and multiple cutoffs as in our motivating example.

%%%%%%%%%%%%%%%%%%%%%%%%%%%%%%%%%%%%%%%%%%%
%%%%%%%%%%%%%%%%%%%%%%%%%%%%%%%%%%%%%%%%%%%
\subsection{Other areas of application}
\label{sec_other_areas}

Besides our motivating example, there are other areas of application where complex treatment rules with multiple (possibly discrete) running variables and multiple cutoffs that lead to the same treatment arise.
%\textcolor{black}{There are other areas where complex treatment rules arise.}
For example, in education, an important question relates to the effects of free-college tuition programs on long-term outcomes. In Chile, free-college programs are in place through specific rules that combine several running variables, namely, four college admission test scores (in language, mathematics, science, and history) in addition to high school Grade Point Average (GPA) and household income. Students with household income below the country's median and test scores and GPA above certain cutoffs are eligible for the free-college tuition program. However, the cutoffs vary from college to college, defining different treatment rules with several running variables across colleges.

%In labor economics, a complex discontinuity design arises in the study of the effects of the Unemployment Insurance (UI) program in Brazil.  In this program, a worker who is laid off receives the UI benefits if two conditions are satisfied: the worker has at least six months of job tenure at layoff and there are at least 16 months between a worker's layoff date and the layoff date of his/her last successful job application.  In this setting, covariates other than the running variables play a central role since unconditional balance checks render standard assumptions for identification implausible; however, it is plausible that, conditional on covariates, treatment assignment is independent of the potential outcomes in a neighborhood of the cutoffs.

In health care policy, the effect of the US Medicare program on health care utilization and patient well-being is a question of great interest. Entry into Medicare is based on several criteria, the most common of which is turning age 65 and having paid taxes for ten years or more. However, one can also enter Medicare through other criteria, including disability and end-stage renal disease.  Any of these eligibility criteria could enable a person to enter the program. End-stage renal disease is in part determined by laboratory tests of kidney function.
%\footnote{\scriptsize{\url{https://www.medicare.gov/manage-your-health/i-have-end-stage-renal-disease-esrd/signing-up-for-medicare-if-you-have-esrd}}}
Disability is determined by physical or mental diagnoses.
%\footnote{\scriptsize{\url{https://www.kidney.org/sites/default/files/docs/ckd_evaluation_classification_stratification.pdf}}}
While many studies
have used the age 65 discontinuity, to our knowledge, no studies have combined that with the additional eligibility criteria.

Finally, in clinical medicine, there are many potential examples where the same treatment or clinical service is indicated for multiple reasons. For example, if one is interested in the effect of a procedure, such as surgery, on an outcome, that procedure could be triggered due to various reasons such as bleeding, infection, or trauma, each with its own threshold. In this and other settings in medicine, where many clinical indications could qualify a patient for a given type of treatment, our framework could potentially be used.

%%%%%%%%%%%%%%%%%%%%%%%%%%%%%%%%%%%%%%%%%%%
%%%%%%%%%%%%%%%%%%%%%%%%%%%%%%%%%%%%%%%%%%%
\subsection{Contribution and outline of the paper}
\label{sec_outline}

Motivated by an observational study of the impact of grade retention on educational and juvenile crime outcomes \textcolor{black}{in Chile}, we propose a framework for complex discontinuity designs, where treatment assignment may be determined by multiple treatment rules, each with multiple running variables and several cutoffs, that lead to the same treatment. 
\textcolor{black}{To our knowledge, existing approaches to discontinuity designs cannot readily accommodate such complex treatment rules. While it may be possible to analyze complex discontinuity designs under existing frameworks, e.g., by collapsing and simplifying the components of a complex treatment rule into a single dimension, this process may overlook important aspects of the assignment mechanism, such that its original structure is not preserved. Moreover, relevant information can be lost, and one may end up comparing treated and control units that are distant from the cutoff on one of the running variables. Fundamentally, the identification assumptions of existing frameworks are not plausible in the context of our case study, because relevant covariates are imbalanced even in the smallest possible neighborhood around the cutoffs.}
In the proposed framework, the observed covariates other than the running variable(s) play a central role for identification of \textcolor{black}{average} treatments effects in a neighborhood of the cutoffs via a local strong ignorability assumption; that is, via local unconfoundedness and local positivity assumptions.
\textcolor{black}{Our work builds on the local randomization framework as formalized by \cite{cattaneo2015randomization} and \cite{li2015evaluating}, where treatment assignment is analyzed as in a local randomized experiment.
More specifically, we follow}
\citet{keele2015enhancing}, who use the local unconfoundedness assumption in the context of a geographic discontinuity design (see also \citealp{keele2017overview}), and \citet{branson2018local}, who extend the local randomization framework and formalize general assignment mechanisms with varying propensity scores. 
Similar assumptions have been invoked by \cite{battistin2008ineligibles}, \cite{angrist2015wanna}, and \cite{forastiere2017selecting}.
\cite{mattei2016regression} articulates how this assumption and positivity form local strong ignorability.
All these frameworks and methods, however, encompass simple treatment rules.

\textcolor{black}{From a methodological standpoint,} one of the points we wish to make in this paper is that, under the assumption of local strong ignorability, we cannot only handle complex discontinuity designs in a straightforward manner, but also facilitate different outcome analyses than those conventionally done in discontinuity designs.
As we illustrate, this framework facilitates simple graphical displays of the outcomes across treatment groups as is done in the analysis of clinical trials, sensitivity analyses on near equivalence tests in matched observational studies, and methods for outcome analysis that combine matching and weighting with regression adjustments.
\textcolor{black}{To our knowledge, none of these analyses have been performed in discontinuity designs and some of them are facilitated by our framework.}
Further, under assumptions that we formalize for the first time in the context of a discontinuity design, we can also generalize the study findings of the discontinuity design to a target population.
\textcolor{black}{The validity of these analyses is predicated in forms of the strong ignorability assumption.}
We note that this assumption will not always be plausible and is different from the one invoked under the continuity-based framework.
The plausibility of these assumptions will depend on their specific context.

\textcolor{black}{From a substantive standpoint, our work contributes to the educational policy literature by providing new evidence of the impact of grade retention on crime and educational outcomes in a developing country. The case of Chile is relevant because its educational system is especially market-oriented and economically segregated, with high levels of grade retention among socially vulnerable students.\footnote{In fact, in 2007 the overall grade retention rate was 5.8\%, with a rate of 6.7\% among students attending public schools (more vulnerable students), as opposed to just 1.7\% among students in private schools.} In the United States, previous works that examine the effect of school grade retention using discontinuity designs include: \cite{jacob2004rd}, who documented that grade retention increased the probability of dropping out of school among students in Chicago; \cite{schwerdt2017rd}, who found that third grade retained students in Florida tended to have higher grade point averages, but no difference in their probability of graduating; and \cite{erena2017test}, who found no effect of grade retention on juvenile crime for 4th grade students and a small negative effect for 8th grade students in Louisiana, while \cite{erena2018test} found that grade retention had a positive impact on 8th graders' future likelihood of committing crime during adulthood. In Chile, \cite{grau2018crime} estimated the effect of grade retention on school dropout and juvenile crime, finding a negative effect on both outcomes. Our work expands on the latter by covering more study cohorts and analyzing a more comprehensive set of educational and criminal outcomes. Furthermore, we use a framework for complex decision rules in order to analyze the effect of grade retention in terms of the primitive rules that determine it.  All previous works follow the continuity-based framework, which is not viable in our case study because its basic assumptions are not satisfied in the observed data.}

%To develop and illustrate this framework, we use as a running example a study of the impact of grade retention on educational and juvenile crime outcomes in Chile.
%In Section \ref{sec_grade}, we explain the school grade retention rules in Chile.
This paper is organized as follows.
In Section \ref{sec_educational}, we describe the data of our case study.
%: a large administrative data set with extensive educational and criminal records of the same students, observed for 15 years.
In Section \ref{sec_framework}, we present our framework for complex discontinuity designs, including the notation, estimands, and assumptions for identification and generalization.
In Section \ref{sec_estimation}, we discuss estimation using matching, weighting, and regression-assisted matching and weighing approaches.
In Section \ref{sec_selection}, we propose methods for selecting the neighborhood for analysis.
In Section \ref{sec_results}, we present the results of our case study.
%different analyses that can be done in discontinuity designs.
In Section \ref{sec_further_extensions}, we outline extensions to the fuzzy case.
In Section \ref{sec_concluding}, we conclude with some remarks.

%% file: dd89_sec2_01.tex
In our case study, we use an administrative data set with extensive educational and criminal records of the same students, followed for 15 consecutive years.
%To study the impact of grade retention on educational and juvenile crime outcomes in Chile, we use an administrative data set with extensive educational and criminal records of the same students, followed for 15 consecutive years.
We assembled this data set from administrative records from the Ministries of Education and Justice in Chile.
The resulting data set covers the period 2002-2016 and includes all students who were enrolled in the Chilean educational system.
The data set contains detailed educational and sociodemographic variables of the students, their families, and their schools.
In particular, between 2002 and 2006, our baseline covariates include the student's gender, age, attendance, grades, and standardized test scores, the parents' education and incomes, and the school and grade attended.
In 2007, we have the school grades of all students, disaggregated by subject.
%As explained in Section \ref{sec_grade}, these variables determine grade retention in Chile, and are central to our discontinuity design.
At the end of 2007, we register whether the student was retained because of low school grades (the school year ends in December in Chile).\footnote{Specifically, we measure grade retention during elementary education, when students generally take the same subjects and retention rules 1 and 2 apply nationally.
In Chile, students are evaluated within schools (for this reason, our outcome comparisons are within schools and school grades, among other covariates, by exact matching; see Section \ref{sec_results} for details).}
%Between 2008 and 2016, we measure educational outcomes of the student and whether he or she was prosecuted for a criminal offense.
Between 2008 and 2016, we measure as outcomes the yearly average school grade in 2008-2011, and the binary variables of grade retention, school dropout and prosecution for a criminal offense in 2008-2016.
In total, we have 1,377,089 student observations in our data set; 4.4\% of them repeated the grade they had taken in 2007, and 3.3\% of them were prosecuted for a criminal offense between 2008 and 2016. 

%% file: dd89_sec3_01.tex
We consider complex discontinuity designs where the treatment assignment is determined by multiple rules that may lead to the same treatment.
Importantly, each rule may depend on several running variables and some running variables may be common to multiple rules.
For each running variable, there may be various cutoffs and neighborhoods around those cutoffs where the identification assumptions \textcolor{black}{hold.
To} describe this \textcolor{black}{framework}, we introduce the following notation.

%%%%%%%%%%%%%%%%%%%%%%%%%%%%%%%%%%%%%%%%%%%
%%%%%%%%%%%%%%%%%%%%%%%%%%%%%%%%%%%%%%%%%%%
\subsection{Notation}

For each unit $i$, let $\boldsymbol{R}_{ij}$ be the vector of running variables that determine treatment assignment rule $j$, with $j = 1, ..., \mathrm{J}$. 
For each $\boldsymbol{R}_{ij}$, let $R_{ijk}$ be its $k$th component, with $k = 1, ..., \mathrm{K}_{\mathrm{j}}$. 
Based on $\boldsymbol{R}_{ij}$, each unit is assigned to treatment $j$ according to the rule
\begin{equation}
	Z_{ij} =
	\begin{cases}
		1, & \text{if} \ R_{ijk} \in (- \infty, c_{jk}) \ \text{for all} \ k = 1, ..., \mathrm{K}_{\mathrm{j}} \\
		0, & \text{otherwise,}
	\end{cases}
\end{equation}
where $c_{jk}$ is the cutoff for running variable $k$ under rule $j$.
Let $Z_i$ denote the treatment indicator, with $Z_i=1$ if unit $i$ receives the treatment and $Z_i=0$ otherwise. 
In our case study, we consider treatment assignment rules that lead to the same treatment so
\begin{equation}
	Z_{i} =
	\begin{cases}
		1, & \text{if} \ Z_{ij}=1 \ \text{for some} \ j = 1, ..., \mathrm{J}
 \\
		0, & \text{otherwise.}
	\end{cases}
\end{equation}

Let $N_{ij}$ be an indicator variable that denotes if unit $i$ is in a neighborhood of the cutoffs of assignment rule $j$. 
Specifically,
\begin{equation}
	N_{ij} =
	\begin{cases}
		1, & \text{if} \ R_{ijk} \in [c_{jk}-\underline{\delta}_{jk}, c_{jk}+ \overline{\delta}_{jk}] \ \text{for all} \ k = 1, ..., \mathrm{K}_{\mathrm{j}} \\
		0, & \text{otherwise,}
	\end{cases}
\end{equation}
\noindent where $\underline{\delta}_{jk}$ and $\overline{\delta}_{jk}$ are positive scalars that define the neighborhood below and above cutoff $jk$, respectively. 
Let $N_{i}$ be an indicator variable that denotes if unit $i$ is in a neighborhood of cutoffs where the identification assumptions (described in Section \ref{sec_assumptions}) hold
\begin{equation}
	N_{i} =
	\begin{cases}
		1, & \text{if} \ N_{ij}=1 \ \text{for some} \ j = 1, ..., \mathrm{J}
 \\
		0, & \text{otherwise.}
	\end{cases}
\end{equation}

\begin{figure}[h!]
\caption{In our case study, there are two treatment rules that lead to the same treatment, which is grade retention. 
Under treatment rule 1, the treatment assignment region is given by the area with dashed lines in the left figure. 
Treatment rule 2 is depicted in the right figure. 
The grey shaded areas denote the neighborhoods where the identification assumptions plausibly hold. 
In general, these areas do not need to be symmetric around the cutoffs.}
\vspace{-.7cm}
\label{fig_notation}
\begin{center}
\begin{subfigure}[b]{0.475\textwidth}
	\includegraphics[width=1\textwidth]{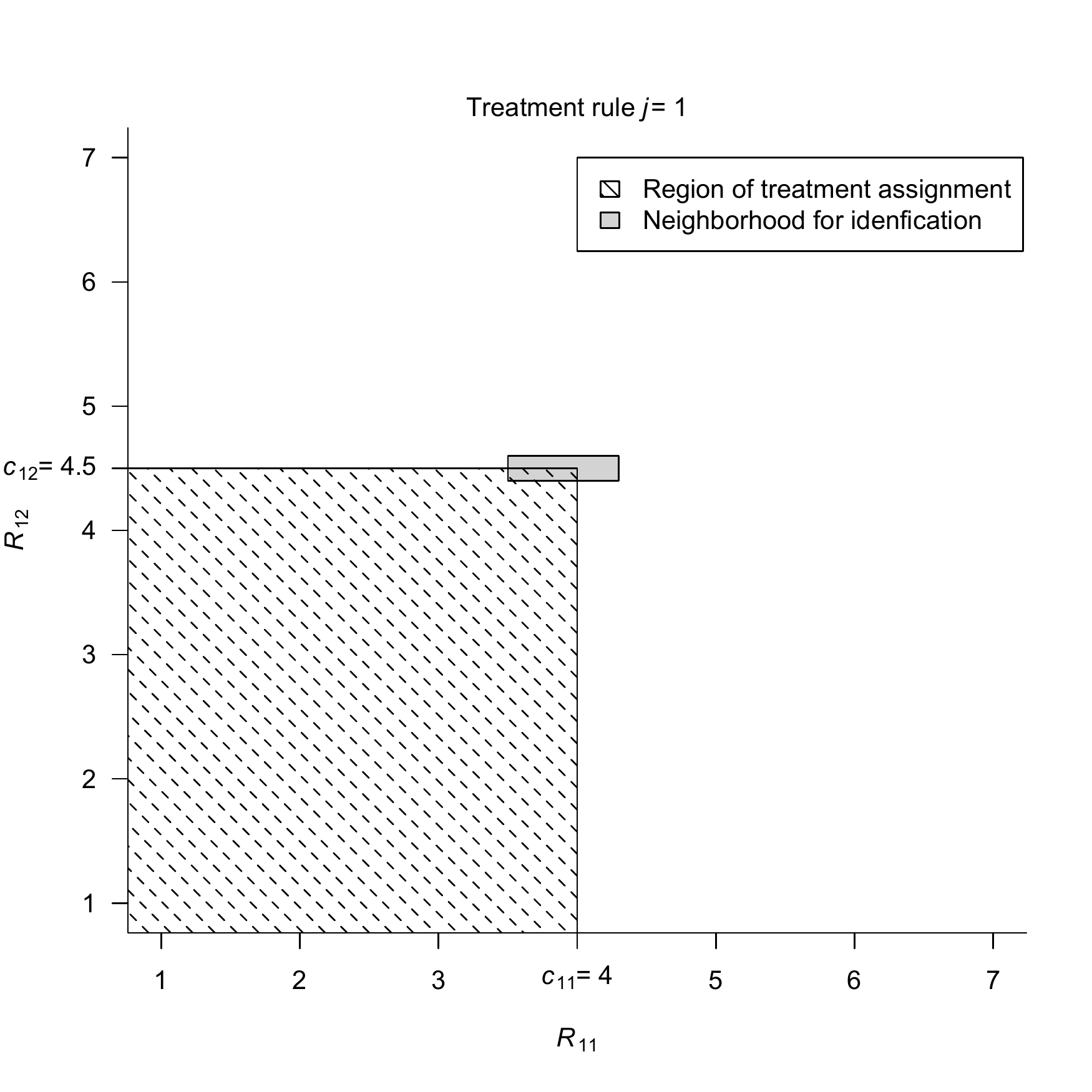}
\end{subfigure}
\begin{subfigure}[b]{0.475\textwidth}
	\includegraphics[width=1\textwidth]{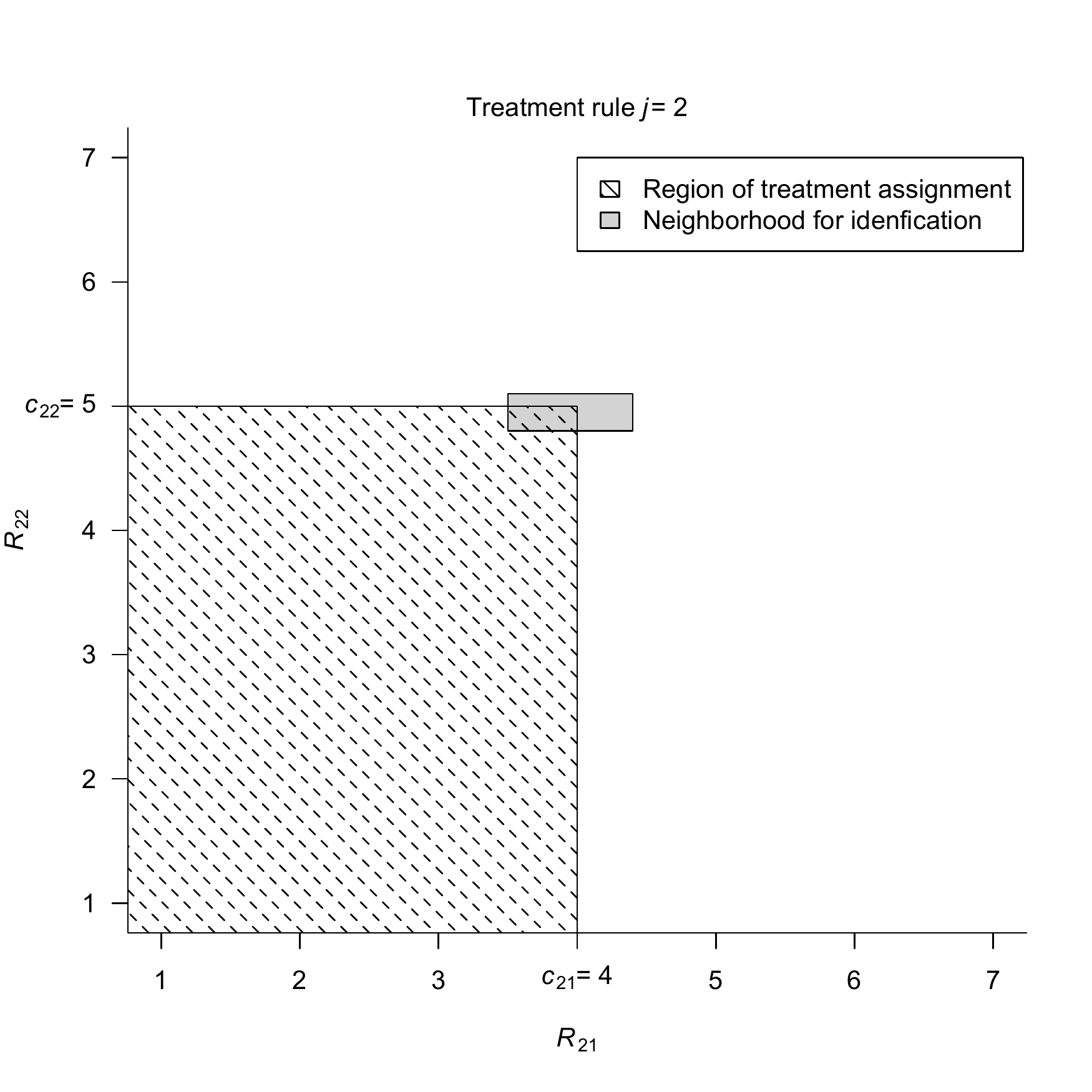}
\end{subfigure}
\end{center}
\vspace{-.75cm}
\end{figure}

By way of illustration, in our case study there are $\mathrm{J} = 2$ treatment rules, with $\boldsymbol{R}_{1}=(R_{11},R_{12})^\top$, where $R_{11}$ is the lowest grade across all school subjects and $R_{12}$ is the average grade across all subjects, and $\boldsymbol{R}_{2}=(R_{21},R_{22})^\top$, where $R_{21}$ is the second lowest grade across all school subjects and $R_{22}$ is the average grade across all subjects, so $R_{12} = R_{22}$. 
Units are in the neighborhood of rule 1 ($N_1 = 1$) if $\boldsymbol{R}_{1} \in [c_{11} - \underline{\delta}_{11}, c_{11} + \overline{\delta}_{11}] \times [c_{12} - \underline{\delta}_{12}, c_{12} + \overline{\delta}_{12}]$, and/or in the neighborhood of rule 2 ($N_2 = 1$) if $\boldsymbol{R}_{2} \in [c_{21} - \underline{\delta}_{21}, c_{21} + \overline{\delta}_{21}] \times [c_{22} - \underline{\delta}_{22}, c_{22} + \overline{\delta}_{22}]$.
\textcolor{black}{Here, a student can be simultaneously in the neighborhood of two cutoffs, for example, of $c_{11}$, corresponding the lowest grade across all school subjects, and of $ c_ {12} = c_ {22} $, the average grade across all subjects.}
See Figure \ref{fig_notation} for details.

%\vspace{-.5cm}

Now, let $Y_i(1)$ and $Y_i(0)$ denote the potential outcomes of unit $i$ under treatment and control, respectively.
%This notation implicitly makes the Stable Unit Treatment Value Assumption (SUTVA, \citealt{rubin1980randomization}), which states that the treatment assignment of one unit does not affect the potential outcomes of other units (i.e., that there is no interference among individuals) and that there are no hidden versions of the treatment that would result in different potential outcomes beyond those encoded by $Y_i(1)$ and $Y_i(0)$. 
This notation implicitly makes the Stable Unit Treatment Value Assumption (SUTVA; \citealp{rubin1980randomization}), which states that the treatment assignment of one unit does not affect the potential outcomes of other units and that there are no hidden versions of the treatment beyond those encoded by the treatment assignment indicator.
In our framework, SUTVA is required only for units in a neighborhood of the cutoffs.
Finally, let $\boldsymbol{X}_i$ denote a vector of observed, pretreatment covariates other than the running variables.

%%%%%%%%%%%%%%%%%%%%%%%%%%%%%%%%%%%%%%%%%%%%
%%%%%%%%%%%%%%%%%%%%%%%%%%%%%%%%%%%%%%%%%%%%
\subsection{Estimand}
\label{sec_estimand}

We wish to estimate the average treatment effect in a neighborhood of the cutoffs where the treatment assignment is unconfounded given the observed covariates (which we term the ``discontinuity neighborhood'').
We call this estimand the Neighborhood Average Treatment Effect (NATE) and define it as
\begin{equation}\label{estimand1}
\tau_{\rm{NATE}} =
\mathbb{E}\big\{Y_i(1)-Y_i(0) \ \big| \ N_i=1\big\}
=\mathbb{E}\bigl[\mathbb{E}\bigl\{Y_i(1)-Y_i(0) \ \big| \ \boldsymbol{X}_i, \ N_i=1\bigl\}\big| N_i=1\bigl].
\end{equation}

In our case study, this is the average effect of grade retention on criminal and educational outcomes for similar students in the discontinuity neighborhood. 
Possibly, this average effect varies across treatment rules; e.g., the effect of repeating under rule 1 differs from rule 2. 
Likewise, the effect of repeating could vary across running variables within the same rule; e.g., the effect of repeating differs across school subjects
%, or the effect of not repeating is different if the student barely passed according to one or both of the school grade running variables that comprise the rule 
(see \citealp{wong2013analyzing}, and \citealp{choi2018regression}, for a related discussion and methods). 
Furthermore, the effects could be modified by covariates; e.g., the effect of repeating is different for boys and girls.
In brief, within the discontinuity neighborhood the effects can be heterogeneous on the treatment rules, running variables, and the covariates. 
%The existence of heterogeneous effects is an empirical question that applies to all studies of causal effects, experimental or observational. 
From (\ref{estimand1}), we note that $\mathbb{E}\bigl\{Y_i(1)-Y_i(0) \ \big| \ \boldsymbol{X}_i, \ N_i=1\bigl\}$ is the Conditional Average Treatment Effect (CATE) in the discontinuity neighborhood (see Chapter 12 of \citealp{imbens2015causal}, for a discussion of the CATE). 
Under the following assumptions, all these heterogenous effects can be analyzed as in observational studies with strongly ignorable treatment assignment, albeit confined to a discontinuity neighborhood.

%%%%%%%%%%%%%%%%%%%%%%%%%%%%%%%%%%%%%%%%%%%%
%%%%%%%%%%%%%%%%%%%%%%%%%%%%%%%%%%%%%%%%%%%%
\subsection{Assumptions}
\label{sec_assumptions}

%%%%%%%%%%%%%%%%%%%%%%%%%%%%%%%%%%%%%%%%%%%
\subsubsection{Assumptions for identification}
\label{sec_assumptions_for_identification}

In this section we state the assumptions needed to identify the NATE.

\begin{assumption}
[local strong ignorability of treatment assignment via the running variables]
~\\*
~\\*
\label{as_local_strong_ignorability}
\; \; \; {\bf Assumption 1a}
\begin{equation*}
\{\boldsymbol{R}_{ij}: j=1,...,\mathrm{J}\} \independent \{Y_i(1), Y_i(0)\} \, | \, \boldsymbol{X}_i, N_i=1\end{equation*}
\; \, \, \, {\bf Assumption 1b}
\begin{equation*}
0<\mathrm{Pr}(Z_{i}=1 \ | \ \boldsymbol{X}_i, N_i=1)<1
\end{equation*}
\end{assumption}

%Assumption \ref{as_local_strong_ignorability} has two components: local unconfoundedness and local positivity of treatment assignment via the running variables.
%Assumption \ref{as_local_strong_ignorability}\textcolor{black}{a} states that, in a neighborhood of the cutoffs, the running variables --- and, therefore, the assignment to treatments --- are independent of the potential outcomes given the observed covariates.
%%One implication of this assumption is that, \textcolor{black}{given} the observed covariates, the regression functions are constant (flat) functions of the running variables in a neighborhood of the cutoffs.
%See \cite{cattaneo2015randomization} for a discussion under the local randomization framework with simple treatment rules.
%Assumption \ref{as_local_strong_ignorability}\textcolor{black}{b} states that, in a neighborhood of the cutoffs, every unit has a positive probability of receiving any version of the treatment given the observed covariates.
%%The assumption of positivity is sometimes called the assumption of common support.
%See \cite{li2015evaluating} for a discussion of this assumption in the context of the local randomization framework.

Assumption \ref{as_local_strong_ignorability} has two components: local unconfoundedness and local positivity of treatment assignment via the running variables.
Assumption \ref{as_local_strong_ignorability}\textcolor{black}{a} states that, in a neighborhood of the cutoffs, the running variables --- and, therefore, the assignment to treatments --- are independent of the potential outcomes given the observed covariates.
Assumption \ref{as_local_strong_ignorability}\textcolor{black}{b} states that in this discontinuity neighborhood every unit has a positive probability of receiving any version of the treatment given the observed covariates.
%The assumption of positivity is sometimes called the assumption of common support.
See \cite{cattaneo2015randomization} and \cite{li2015evaluating} for a discussion of these assumptions under the local randomization framework with simple treatment rules.

Under Assumptions \ref{as_local_strong_ignorability}\textcolor{black}{a} and \ref{as_local_strong_ignorability}\textcolor{black}{b}, we can non-parametrically identify $\tau_{\rm{NATE}}$ in (\ref{estimand1}) using the following identity:
\begin{align*}\label{identificationestimand2}
\mathbb{E} \big\{ Y_i(1)-Y_i(0) \big| N_i=1\big\}
= \mathbb{E} \big\{
& \mathbb{E}\big( Y_i\ \big| \ Z_i=1, \ \boldsymbol{X}_i, \ N_i=1 \big) \\
& -\mathbb{E}\big( Y_i\ \big| Z_i=0, \ \boldsymbol{X}_i, \ N_i=1 \big) \big| N_i=1\big\}.
\end{align*}

%In essence, Assumptions \ref{as_local_strong_ignorability}\textcolor{black}{a} and \ref{as_local_strong_ignorability}\textcolor{black}{b} are the usual conditions required by the strong ignorability assumption in observational studies \citep{rosenbaum1983central}, but in a neighborhood of the cutoffs.
%In other words, in a neighborhood of the cutoffs, conditional on the observed covariates, a discontinuity design can be analyzed as an observational study under strong ignorability.
In other words, under Assumptions \ref{as_local_strong_ignorability}\textcolor{black}{a} and \ref{as_local_strong_ignorability}\textcolor{black}{b}, a discontinuity design can be analyzed as an observational study under strong ignorability \citep{rosenbaum1983central}, but confined to a discontinuity neighborhood. 
As we discuss below, this ``local strong ignorability'' assumption facilitates a series of tasks that are typically precluded in discontinuity designs under usual assumptions such as handling multiple treatment rules and generalizing effect estimates.
It also suggests using other methods for outcome analyses normally used in observational studies such as those described in Section \ref{sec_estimation}.
For a more detailed discussion about the assumption of strong ignorability in observational studies, see, e.g., Chapter 12 of \cite{imbens2015causal}.

\textcolor{black}{Assumptions} \ref{as_local_strong_ignorability}\textcolor{black}{a} and \ref{as_local_strong_ignorability}\textcolor{black}{b} are different from the usual assumptions in the continuity-based and the local randomization frameworks.
One could argue that the above assumptions are stronger than the typical ones in the continuity-based framework;\footnote{In the continuity-based framework the mean potential outcome functions need to be continuous functions of the running variable whereas our framework requires them to be constant given the observed covariates.} however, the frameworks target different estimands (see, for instance, \citealp{de2016misunderstandings}, and \citealp{mattei2016regression}).
In our framework the estimand is the Neighborhood Average Treatment Effect, whereas in the continuity-based framework the estimand is the average effect \emph{at} the cutoff.
In both cases, these are local estimands (as opposed to estimands for the general population), although the local randomization framework can facilitate the generalization of causal inferences to other populations (see Section \ref{sec_assumptions_for_generalization}).
%\textcolor{black}{In both cases, these are local estimands (as opposed to estimands for the general population), although the generalization of causal inferences can be more easily conceptualized under the local randomization framework (see Section \ref{sec_assumptions_for_generalization}).}
%\textcolor{black}{In both cases, these are local estimands (as opposed to estimands for the general population), yet under the local randomization framework is straightforward to generalize inferences (see Section \ref{sec_assumptions_for_generalization}).}
We can say that there is a trade-off between invoking different assumptions and identifying different estimands.
While our framework builds on the local randomization framework, its assumptions are more plausible in practice than the ones in typical local randomization settings, because in many applications local independence conditional on covariates is a more realistic assumption than unconditional local independence.
In our case study, even in the smallest possible discontinuity neighborhood there are imbalances in relevant covariates measuring the student's ability. However, after adjusting for these and other educational and sociodemographic covariates, treatment assignment can be considered as-if randomized in a discontinuity neighborhood, because other covariates are balanced, but not across the entire range of the running variables.
\textcolor{black}{
%The idea is that conditional on observed educational and sociodemographic characteristics of the students, their families, and their schools that affect grade retention in Chile, retention is essentially random in a discontinuity neighborhood but not outside of it. 
The validity of these assumptions is, of course, context-specific and must be carefully evaluated in the setting under study.}\footnote{\textcolor{black}{In our case study, the assumptions of the continuity-based framework are not satisfied because relevant covariates are imbalanced even in the smallest possible neighborhood around the cutoffs (see Appendix H in the Supplementary Materials).}
The local randomization implies that the mean potential outcome functions do not depend on the running variables $R_{11}$ (the lowest grade across all school subjects), $R_{21}$ (the second lowest grade), and $R_{12} = R_{22}$ (the average grade across all subjects) in the discontinuity neighborhood.
However, more skillful students, who are likely to obtain higher grades in these variables, are also likely to be systematically different from those students whose scores are lower (less skillful students).
Therefore, it is not plausible that the mean potential outcome functions are constant functions of the grades, even in the discontinuity neighborhood.
A more plausible assumption is that the mean potential outcome functions are constant functions of the grades after adjusting for covariates.
This is stated by assumptions \ref{as_local_strong_ignorability}\textcolor{black}{a} and \ref{as_local_strong_ignorability}\textcolor{black}{b}.
In other words, if we are able to adjust for covariates that capture the students' skills and the environment where they are educated (such as their standardized test scores before exposure to the treatment, their previous educational attainments, school attended, and teachers' characteristics), then it is more plausible that the mean potential outcomes do not depend on the running variables in the discontinuity neighborhood.
As we describe below, a feature of this framework is that assumptions \ref{as_local_strong_ignorability}\textcolor{black}{a} and \ref{as_local_strong_ignorability}\textcolor{black}{b} in principle have testable implications, which are useful to select the neighborhood for analysis.}
%Finally, this framework naturally handles cases where the running variables are discrete and intermediate outcomes are present.

%%%%%%%%%%%%%%%%%%%%%%%%%%%%%%%%%%%%%%%%%%%
\subsubsection{Assumptions for generalization}
\label{sec_assumptions_for_generalization}

For generalization --- that is, for identification and estimation of average treatment effects in target populations beyond the sample considered in the analysis --- we consider two relevant cases.
In both cases, selection into the sample is determined by observed covariates, in addition to the running variables.
The key distinction is whether the target population has values of the running variables inside the neighborhood.
The case where the target population has values of the running variables outside the neighborhood naturally involves stronger assumptions that, as we discuss below, contradict Assumptions \ref{as_local_strong_ignorability}\textcolor{black}{a} and \ref{as_local_strong_ignorability}\textcolor{black}{b} because they require treatment assignment to be strongly ignorable throughout the \textcolor{black}{entire range} of the running variables.

Let $\mathscr{P}$ be a target population and $\mathcal{S}$ be the sample of units from $\mathscr{P}$ selected into the neighborhood of the cutoffs where assumptions \ref{as_local_strong_ignorability}\textcolor{black}{a} and \ref{as_local_strong_ignorability}\textcolor{black}{b} \textcolor{black}{are believed to hold.}
Write $S_i = 1$ if unit $i$ is selected into the sample and $S_i = 0$ otherwise.
Here, we wish to identify and estimate the target average treatment effect (TATE; \citealt{kern2016assessing})
\begin{equation*}
\tau_{\textrm{TATE}} :=  \mathbb{E}_\mathscr{P} \{ Y_i(1)-Y_i(0) \},
%\tau_{\textrm{TATE}} := \frac{1}{|\mathscr{P}|} \sum_{i \in \mathscr{P}} Y_i(1) - Y_i(0).
\end{equation*}
\noindent that is, the average treatment effect in the target population $\mathscr{P}$.
%; if $\mathscr{P}$ is finite, $\tau_{\textrm{TATE}} = \frac{1}{|\mathscr{P}|} \sum_{i \in \mathscr{P}} Y_i(1) - Y_i(0)$.

%%%%%%%%%%%%%%%%%%%%%%%%%%%%%%%%%%%%%%%%%%%
\subsubsection*{$\bullet$ Case 1: $\mathscr{P}$ such that $N_i=1$ for all $i \in \mathscr{P}$} 
In this case, we assume the following two conditions hold.

\begin{assumption}
[strong ignorability of study selection in a neighborhood]
~\\*
~\\*
\label{as_local_strong_ignorability_sel2}
\; \; \; {\bf Assumption 2a}
\begin{equation*}
\{ Y_i(1), Y_i(0) \} \independent S_i \, | \, \boldsymbol{X}_i, N_i=1
\end{equation*}
\; \, \, \,  {\bf Assumption 2b}
\begin{equation*}
0 < \Pr ( S_i = 1 | \boldsymbol{X}_i, N_i=1 ) < 1
\end{equation*}
\end{assumption}

\vspace{-.75cm}
Assumption \ref{as_local_strong_ignorability_sel2}\textcolor{black}{a} states that selection into the sample is independent of the potential outcomes given the observed covariates and that the running variables take values in the neighborhood of the cutoffs.\footnote{This assumption can be relaxed to require conditional mean independence only; that is, $\mathbb{E} \{ Y_i(z) | \boldsymbol{X}_i, S_i = 1, N_i=1 \} = \mathbb{E} \{ Y_i(z) | \boldsymbol{X}_i, N_i=1 \}$.}
Assumption \ref{as_local_strong_ignorability_sel2}\textcolor{black}{b} states that every unit in the target population $\mathscr{P}$ has a positive probability of selection into the sample given identical conditions on the observed covariates and running variables.

Under assumptions \ref{as_local_strong_ignorability}\textcolor{black}{a}, \ref{as_local_strong_ignorability}\textcolor{black}{b}, \ref{as_local_strong_ignorability_sel2}\textcolor{black}{a}, and \ref{as_local_strong_ignorability_sel2}\textcolor{black}{b}, we can identify $\tau_{\textrm{TATE}}$ \textcolor{black}{from}
\begin{equation*}
\mathbb{E}_\mathscr{P}\big\{Y_i(z)\big| N_i=1\big\} =\mathbb{E}_\mathscr{P}\big\{
\mathbb{E}\big(Y_i \big| S_i=1, \ Z_i=z, \ \boldsymbol{X}_i, \ N_i=1\big)\big| \ N_i=1\big\},
\end{equation*}
\noindent for $z = 0, 1$.

%%%%%%%%%%%%%%%%%%%%%%%%%%%%%%%%%%%%%%%%%%%
\subsubsection*{$\bullet$ Case 2: $\mathscr{P}$ such that $N_i\neq1$ for some $i \in \mathscr{P}$} 
%\vspace{-.5cm} 

In this case, we assume the following two conditions hold.

\begin{assumption}
[strong ignorability of study selection]
~\\*
~\\*
\label{as_3}
\; \; \; {\bf Assumption 3a}
\begin{equation*}
\{ Y_i(1), Y_i(0) \} \independent S_i \, | \, \boldsymbol{X}_i
\end{equation*}
\; \, \, \,  {\bf Assumption 3b}
\begin{equation*}
0 < \Pr ( S_i = 1 | \boldsymbol{X}_i ) < 1
\end{equation*}
\end{assumption}

\vspace{-.75cm}
Assumptions \ref{as_3}\textcolor{black}{a} and \ref{as_3}\textcolor{black}{b} are stronger than Assumptions \ref{as_local_strong_ignorability_sel2}\textcolor{black}{a} and \ref{as_local_strong_ignorability_sel2}\textcolor{black}{b}.
%\textcolor{red}{because they require the study selection mechanism to be conditionally independent of the potential outcomes for all units as opposed to units in a neighborhood only}.
Under these assumptions, we need to modify assumptions \ref{as_local_strong_ignorability}\textcolor{black}{a} and \ref{as_local_strong_ignorability}\textcolor{black}{b} in order to identify $\tau_{\textrm{TATE}}$.
%In a sense, assumptions \ref{as_3}\textcolor{black}{a} and \ref{as_3}\textcolor{black}{b} contradict assumptions \ref{as_local_strong_ignorability}\textcolor{black}{a} and \ref{as_local_strong_ignorability}\textcolor{black}{b}.
In fact, under these assumptions
%in order to identify $\tau_{\textrm{TATE}}$ from the observed data under assumptions \ref{as_3}\textcolor{black}{a} and \ref{as_3}\textcolor{black}{b}, 
we must have strong ignorability of treatment assignment via the running variables; that is, $\{\boldsymbol{R}_{ij}: j=1,...,\mathrm{J}\} \independent \{Y_i(1), Y_i(0)\} \, | \, \boldsymbol{X}_i$ and $0<\mathrm{Pr}(Z_{i}=1 | \boldsymbol{X}_i)<1$.
In other words, the assignment of the running variables must be strongly ignorable given the observed covariates throughout the entire range of the running variables.
Clearly, these are strong assumptions that deviate from traditional discontinuity settings.

%\begin{assumption}
%[strong ignorability of treatment assignment via the running variable]
%~\\*
%~\\*
%\label{as_4}
%\; \; \; {\bf Assumption 4a}
%\begin{equation*}
%\{\boldsymbol{R}_{ij}: j=1,...,\mathrm{J}\} \independent \{Y_i(1), Y_i(0)\} \, | \, \boldsymbol{X}_i
%\end{equation*}
%\; \, \, \, {\bf Assumption 4b}
%\begin{equation*}
%0<\mathrm{Pr}(Z_{i}=1 | \boldsymbol{X}_i)<1
%\end{equation*}
%\end{assumption}
%
%\vspace{-.75cm}
%In other words, assumptions \ref{as_4}\textcolor{black}{a} and \ref{as_4}\textcolor{black}{b} must hold; that is, the assignment of the running variables must be strongly ignorable given the observed covariates throughout the entire range of the running variables.

%% file: dd89_sec4_01.tex
%We consider three estimation approaches based on matching, weighting, and a combination of the two with additional regression adjustments.

%%%%%%%%%%%%%%%%%%%%%%%%%%%%%%%%%%%%%%%%%%%
%%%%%%%%%%%%%%%%%%%%%%%%%%%%%%%%%%%%%%%%%%%
\subsection{Matching approaches}

Let $\mathcal{I}_T$ and $\mathcal{I}_C$ be the set of indices of the treated and control units in $\mathcal{S}$.
Specifically, let $\mathcal{I}_T = \{ i : \ Z_i=1 \ \textrm{and} \ N_i=1 \}$ and $\mathcal{I}_C = \{ i : \ Z_i=0 \ \textrm{and} \ N_i=1  \}$.
Following \cite{zubizarreta2014matching}, we may find the largest pair-matched sample $\boldsymbol{m}^*$ of treatment and control units that is balanced in a neighborhood of the cutoffs as

\begin{align}
\argmax_{\boldsymbol{m}} \grande\{  & \sum_{t \in \mathcal{I}_T} \sum_{c \in \mathcal{I}_C}  m_{tc} : \label{eq:matching1} \\
& \left| \sum_{t \in \mathcal{I}_T} \sum_{c \in \mathcal{I}_C} m_{tc} \{ B_q(\boldsymbol{X}_t) - \sum_{i \in \mathcal{P}} B_q(\boldsymbol{X}_i) \} \right| \leq \delta_q \sum_{t \in \mathcal{I}_T} \sum_{c \in \mathcal{I}_C} m_{tc}, \: q = 1, \ldots, \mathrm{Q}; \label{eq:matching2} \\
& \left| \sum_{t \in \mathcal{I}_T} \sum_{c \in \mathcal{I}_C} m_{tc} \{ B_q(\boldsymbol{X}_c) - \sum_{i \in \mathcal{P}} B_q(\boldsymbol{X}_i) \} \right| \leq \delta_q \sum_{t \in \mathcal{I}_T} \sum_{c \in \mathcal{I}_C} m_{tc}, \: q = 1, \ldots, \mathrm{Q}; \label{eq:matching3} \\
& m_{tc} \in \{0, 1\}, t \in \mathcal{I}_T, c \in \mathcal{I}_C \label{eq:matching4}
\grande\}
\end{align}
where $m_{tc}$ are binary decision variables that determine whether treated unit $t$ is matched to control unit $c$;
$B_q(\boldsymbol{X}_t)$ are suitable transformations of the observed covariates that span a certain function space;
and $\delta_q$ is a tolerance that restrains the imbalances in the functions $B_q(\cdot)$ of the covariates (see \citealp{wang2020minimal}, for a tuning algorithm).
Thus, the matching given by (\ref{eq:matching1})-(\ref{eq:matching4}) is the maximal size pair-matching that approximately balances the transformations $B_q(\cdot)$ of the covariates relative to the target population $\mathscr{P}$.
Following \cite{zubizarreta2014matching}, we can re-match the matching $\boldsymbol{m}^*$ in order to minimize the total sum of covariate distances between matched units, while preserving aggregate covariate balance.
After matching, we can do inference using randomization techniques as in \cite{rosenbaum2002observational} or the large sample approximations in \cite{abadie2006large}.

%%%%%%%%%%%%%%%%%%%%%%%%%%%%%%%%%%%%%%%%%%%
%%%%%%%%%%%%%%%%%%%%%%%%%%%%%%%%%%%%%%%%%%%
\subsection{Weighting approaches}

%Following \cite{wang2019large}, 
The previous matching approach (\ref{eq:matching1})-(\ref{eq:matching4}) can be seen as a form of weighting where the weights are binary (or a constant fraction) that encode an assignment between matched units.
An alternative is to relax this matching problem into a weighting problem as follows,
\begin{align}
\argmin_{\boldsymbol{w}} \grande\{  & \sum_{i \in \mathcal{I}_C} \left( w_i - \bar{w} \right)^2 : \\
& \left|\sum_{i \in \mathcal{I}_C} w_i Z_i B_q(\boldsymbol{X}_i) - \frac{1}{|\mathcal{I}_T|}\sum_{i \in \mathcal{I}_T} B_q(\boldsymbol{X}_i) \right| \leq \delta_q, \: q = 1, \ldots, \mathrm{Q}; \\
& \sum_{i \in \mathcal{I}_C} w_i = 1, w_i \geq  0, i \in \mathcal{I}_C
\grande\}.
\end{align}

These are the weights $\boldsymbol{w}^*$ of minimum variance that approximately balance the covariates.
The asymptotic properties and inferential methods with balancing weights are discussed in \citet{zhao2019covariate} and \citet{wang2020minimal}.

%%%%%%%%%%%%%%%%%%%%%%%%%%%%%%%%%%%%%%%%%%%
%%%%%%%%%%%%%%%%%%%%%%%%%%%%%%%%%%%%%%%%%%%
\subsection{Regression-assisted approaches}

The above matching and weighting approaches can be \textcolor{black}{complemented with additional regression adjustments}.
Here we present two regression-assisted approaches that follow the above matching and weighting approaches.

For matching, following \cite{rubin1979using}, we may use a Regression Assisted Matching (RAM) estimator of the form
\begin{equation}
\hat{\tau}_{\textrm{RAM}} = (\bar{y}_{T\cdot} - \bar{y}_{C\cdot}) - (\bar{\boldsymbol{x}}_{T\cdot} - \bar{\boldsymbol{x}}_{C\cdot}) \hat{\boldsymbol{\beta}}
\end{equation}
where $\bar{y}_{T\cdot}$ and $\bar{y}_{C\cdot}$ are the means of $Y$ in the matched treated and control samples, and analogously, $\boldsymbol{x}_{T\cdot}$ and $\bar{\boldsymbol{x}}_{C\cdot}$ are the means of $\boldsymbol{x}$.
The vector $\hat{\boldsymbol{\beta}}$ comprises the estimated linear regression coefficients of the matched-pair differences in outcomes on the matched-pair differences in covariates.

For weighting, following \cite{athey2018approximate}, we may use a Regression Assisted Weighting (RAW) estimator of the form
\begin{equation}
\label{eq_raw}
\hat{\tau}_{\textrm{RAW}} = \bar{\boldsymbol{x}}_{T\cdot} \hat{\boldsymbol{\beta}}_{T} - \left\{ \bar{\boldsymbol{x}}_{T\cdot} \hat{\boldsymbol{\beta}}_{C} + \sum_{i \in \mathcal{I}_C} w_i (y_i - \boldsymbol{x}_i \hat{\boldsymbol{\beta}}_{C} ) \right\}
\end{equation}
where $\hat{\boldsymbol{\beta}}_{T}$ and $\hat{\boldsymbol{\beta}}_{C}$ are the estimated regression coefficients for the treated and control samples, respectively.
For inference, given the selected neighborhood, we may use the methods in \citet{athey2018approximate} and \citet{hirshberg2018augmented}.
See \citet{robins1994estimation} and \citet{abadie2011bias} for \textcolor{black}{related estimators}.

%% file: dd89_sec5_01.tex
A central question in practice is how to select a neighborhood for analysis; that is, the region around the cutoffs where the identification assumptions are likely to hold.
Assumptions \ref{as_local_strong_ignorability}\textcolor{black}{a} and \ref{as_local_strong_ignorability}\textcolor{black}{b} require that at least one neighborhood exists.
In order to maximize the precision of the study, we will find the largest neighborhood where the assumptions are plausible.

Assumption \ref{as_local_strong_ignorability}\textcolor{black}{a} is not directly testable from the observed data.
%The reason is that, for each unit, we can observe only one of the potential outcomes.
However, there are implications of this assumption that can be tested.
In this paper, we present two sets of methods for testing such implications and selecting the neighborhood.
The first set of methods is design-based in the sense that they do not use outcome information.
%; instead, they use auxiliary sets of covariates other than the ones needed for conditioning on assumptions \ref{as_local_strong_ignorability}\textcolor{black}{a} and \ref{as_local_strong_ignorability}\textcolor{black}{b}.
% (either secondary covariates, lagged outcomes, or lagged running variables).
The second set of methods is semi-design-based in that they use outcome information, but in a planning sample, separate from the sample used for the actual outcome analyses.
We present this second set of methods in Appendix G in the Supplementary Materials.\footnote{See \cite{cattaneo2015randomization}, \cite{li2015evaluating}, and \cite{cattaneo2016choice} for methods for selecting a neighborhood in discontinuity designs in the local randomization framework.
Also, see \cite{branson2018local} for related methods in the setting of local randomization given covariates with general assignment mechanisms under simple treatment rules.}

%%%%%%%%%%%%%%%%%%%%%%%%%%%%%%%%%%%%%%%%%%%%
%%%%%%%%%%%%%%%%%%%%%%%%%%%%%%%%%%%%%%%%%%%%
%\subsection{Design-based approaches}
%\label{sec_design}

%Assumption \ref{as_local_strong_ignorability}\textcolor{black}{a} says that treatment assignment is independent of the potential outcomes conditional on the observed covariates in a neighborhood of the cutoffs.
One implication of \ref{as_local_strong_ignorability}\textcolor{black}{a} is balance on the potential outcomes and any other variable measured before treatment assignment, given the observed covariates $\boldsymbol{X}$.
Let $\boldsymbol{X}^{\textrm{test}}$ denote such other variables.
These variables can be secondary covariates, lagged outcomes, or lagged running variables, and are different from the observed covariates $\boldsymbol{X}$ required in Assumption \ref{as_local_strong_ignorability}\textcolor{black}{a}.
Conditional on $\boldsymbol{X}$, Assumption \ref{as_local_strong_ignorability}\textcolor{black}{a} implies that $\boldsymbol{X}^{\textrm{test}}$ is balanced across treatment groups in terms of its joint distribution, and therefore, in terms of its moments.

To assess Assumption \ref{as_local_strong_ignorability}\textcolor{black}{a}, we may test for joint, multivariate covariate balance on $\boldsymbol{X}^{\textrm{test}}$ conditional on $\boldsymbol{X}$.
For instance, we may use the cross-match test to compare multivariate distributions (\citealp{rosenbaum2005exact, heller2010using}).
%To assess the plausibility of Assumption \ref{as_local_strong_ignorability}\textcolor{black}{a} using $\boldsymbol{X}_i^{\textrm{test}}$, we may test balance in covariates $\boldsymbol{X}_i^{\textrm{test}}$, for instance, using the cross-match test (\citealt{rosenbaum2005exact, heller2010using}).
%The cross-match test is a distribution-free, exact test used to compare multivariate distributions.
%With this test, we will consider Assumption \ref{as_local_strong_ignorability}\textcolor{black}{a} to be plausible if we fail to reject the null hypothesis that $\boldsymbol{X}_i^{\textrm{test}}$ has the same multivariate distribution across treatment groups after adjusting for $\boldsymbol{X}_i$.
We will consider the assumption of local unconfoundedness to be plausible if we fail to reject the null hypothesis that $\boldsymbol{X}^{\textrm{test}}$ has the same multivariate distribution across treatment groups after adjusting for $\boldsymbol{X}$.
In addition, we may test the validity of Assumption \ref{as_local_strong_ignorability}\textcolor{black}{a} by checking the following balance conditions
%\begin{align}\label{balance}
%\mathbb{E}\bigl[\mathbb{E}\bigl\{h(\boldsymbol{X}_i^{\textrm{test}})\ \big|  Z_i=1, \ \boldsymbol{X}_i, \ N_i=1\bigl\} 
%& -\mathbb{E}\bigl\{h(\boldsymbol{X}_i^{\textrm{test}})\ \big| Z_i=0, \ \boldsymbol{X}_i, \ N_i=1\bigl\}\big| N_i=1\bigl]=0.
%\end{align}
\vspace{-.25cm}
\begin{widerequation}\label{balance}
\mathbb{E}\bigl[\mathbb{E}\bigl\{h(\boldsymbol{X}_i^{\textrm{test}})\ \big|  Z_i=1, \ \boldsymbol{X}_i, \ N_i=1\bigl\}  -\mathbb{E}\bigl\{h(\boldsymbol{X}_i^{\textrm{test}})\ \big| Z_i=0, \ \boldsymbol{X}_i, \ N_i=1\bigl\}\big| N_i=1\bigl]=0,
\end{widerequation}
for any function $h(\cdot)$.
For instance, (\ref{balance}) can be tested after matching using randomization techniques.
The idea is to find the largest neighborhood of the cutoffs such that $\boldsymbol{X}^{\textrm{test}}$ is balanced after adjusting for $\boldsymbol{X}$ by matching.
In practice, a conservative way to select this neighborhood is to implement different tests for univariate and multivariate covariate balance and select the largest neighborhood where we fail to reject the null hypothesis of covariate balance for any of the tests.
%See Section \ref{sec_finding} for an example.
See Appendix G for details.

%% file: dd89_sec6_01.tex
\subsection{Finding a neighborhood for analysis}
\label{sec_finding}

To find the discontinuity neighborhood, we follow the design-based approach described in Section \ref{sec_selection}, where secondary covariates are available.
Here, $\boldsymbol{X}^{\textrm{test}}$ is a three-dimensional vector that includes both parents' education (measured in years of schooling), and the household per capita income (measured in pesos).
$\boldsymbol{X}$ includes the student's school attended and grade level in 2007, gender, birth year and month, number of times of previous grade retention, and standardized test scores in language and math.

Under Assumption \ref{as_local_strong_ignorability}\textcolor{black}{a}, $\boldsymbol{X}^{\textrm{test}}$ should be balanced across treatment groups in the discontinuity neighborhood ($N=1$), after conditioning on $\boldsymbol{X}$.
%The idea is that, after conditioning on $\boldsymbol{X}$, if the running variables (and therefore, the treatment) do not affect the potential outcomes in the neighborhood, then nor do they affect the secondary pretreatment variables.
%Since $\boldsymbol{X}^{\textrm{test}}$ is observed irrespective of the realized treatment status, the validity of Assumption \ref{as_local_strong_ignorability}\textcolor{black}{a} can be evaluated using tests for multivariate covariate balance on $\boldsymbol{X}^{\textrm{test}}$ (e.g., the cross-match test).
%In addition, w
We can check the following balance conditions
\begin{eqnarray}\label{balance_case}
  \mathbb{E}(\boldsymbol{X}_i^{\textrm{test}}\ | \ Z_i=1, \ \boldsymbol{X}_i, \ N_i=1)=\mathbb{E}(\boldsymbol{X}_i^{\textrm{test}}\ | Z_i=0, \ \boldsymbol{X}_i, \ N_i=1).
\end{eqnarray}

\indent We use matching to adjust for the observed covariates and select the largest neighborhood that satisfies these conditions.
In the spirit of \cite{cattaneo2015randomization}, we implement a sequence of balance tests on $\boldsymbol{X}^{\textrm{test}}$ conditioning on $\boldsymbol{X}$ for a sequence of expanding neighborhoods and select the largest one where we fail to reject the above balance tests.
We take the following steps in order to select a neighborhood for analysis.

%\onehalfspacing
\begin{enumerate}
\item Start with a small neighborhood around the cutoffs.
\item Find the largest matched sample in the neighborhood that balances the covariates $\boldsymbol{X}$.
\item Test that $\boldsymbol{X}^{\textrm{test}}$ has the same distribution across treatment groups using the cross-match test for multivariate balance and the permutational t-test for mean balance.
	\begin{enumerate}
	\item If the minimum $p$-value is less than the critical value $p^*$, then repeat steps 1 to 3 starting with a smaller neighborhood.
	\item If the minimum $p$-value is greater than or equal to $p^*$, then expand the neighborhood and repeat steps 2 and 3 until the new minimum $p$-value is smaller than $p^*$.
	\end{enumerate}
\item Retain the largest neighborhood with a minimum $p$-value greater than or equal to $p^*$.	
\end{enumerate}
%\doublespacing

In our case study, we set $p^*$ to 0.1 \citep{cattaneo2015randomization}. 
\textcolor{black}{As $p^*$ increases, the balance criteria becomes more stringent, and the resulting neighborhood is smaller; we set $p^*$ to 0.1 because it is is more conservative than setting it to 0.05.}
In step 2 we match students exactly on age (in months), gender, school, school grade, and past grade retention (number of times).
In addition, we match for mean balance for the standardized test scores in language and mathematics.
More specifically, we use cardinality matching (\citealp{zubizarreta2014matching}) to find the largest pair-matched sample that is balanced according to these two exact matching and mean balance requirements.\footnote{Despite the potential cost of dropping a large number of students from the analysis, we view exact matching on age, gender, school, school grade, and past grade retention of primary importance, as it makes Assumption \ref{as_local_strong_ignorability}\textcolor{black}{a} more plausible.
In fact, by matching exactly on covariates, we are also balancing unobserved covariates that are constant within interactions of categories of the exact matching covariates.
For example, by matching exactly on school and school grade, we are also exactly balancing unobserved school and classroom characteristics (including characteristics of the principal and teachers, the school's support network, and the social environment where the school is located) which may vary by age and gender.}
See Appendix A for a summary of covariate balance before and after matching.

%\vspace{-.5cm}
Table $\ref{table:table2}$ summarizes the process of selecting a neighborhood in our case study.
In the first row of the table, we start with a neighborhood defined by $\underline{\delta}_{qr} = \overline{\delta}_{qr} =$ 0.1 for all the cutoffs $qr$ (0.1 is the minimum increment of school grades in Chile).
This results in a matched sample of 243 pairs comprising 486 students.
For this matched sample, we fail to reject the null hypothesis of no differences in secondary covariates (the minimum $p$-value is 0.33); therefore we expand the neighborhood as noted by the second line of the table.
%A natural question to ask is how to expand the neighborhood.
%In typical discontinuity designs with only one running variable, the neighborhood is expanded symmetrically in both directions of the cutoff, but this is a simplification and the neighborhood may be expanded asymmetrically in one direction first.
%In more complex designs with many running variables and multiple treatment rules, like in our running example, there are more directions in which the neighborhood can be expanded first (in our running example, there are eight such directions).
%In our running example, it is more plausible to believe that, after controlling for the observed covariates, students will be comparable in a larger neighborhood of the cutoff of a single school subject ($R_{11}$ or $R_{21}$) than in a neighborhood of equal size of the average across all subjects ($R_{12}$ and $R_{22}$).
%For this reason, we first expand the neighborhoods of $R_{11}$ and $R_{21}$.
%For $R_{11}$ and $R_{21}$ we expand the neighborhood symmetrically for simplicity, but this does not need to be the case in general.
After expanding the neighborhood each time by 0.1 for $R_{11}$ and $R_{21}$ symmetrically (see Appendix G for details), we obtain a matched sample of 1,141 pairs comprising 2,282 students, with a neighborhood of rule 1 given by [3.5, 4.4]$\times$ [4.3, 4.6] and a neighborhood of rule 2 given by [3.5, 4.4] $\times$ [4.8, 5.1].
%, where we fail to reject the null hypothesis of no differences on secondary covariates (minimum $p$-value of 0.16), but 
For the subsequent larger neighborhood, we reject the null hypothesis of balance on secondary covariates (the minimum $p$-value is 0.03). 
%Thus, we retain as our final neighborhood the matched sample with 1,141 pairs of students.
In our notation, the selected neighborhood of rule 1 is  $[c_{11} - \underline{\delta}_{11}, c_{11} + \overline{\delta}_{11}] \times [c_{12} - \underline{\delta}_{12}, c_{12} + \overline{\delta}_{12}]$, and the corresponding neighborhood of rule 2 is $[c_{21} - \underline{\delta}_{21}, c_{21} + \overline{\delta}_{21}] \times [c_{22} - \underline{\delta}_{22}, c_{22} + \overline{\delta}_{22}]$, where $c_{11}=$3.9 and $c_{12}=$4.4 are the cutoff values of $R_{11}$ and $R_{12}$ under the rule 1, $c_{21}=$3.9 and $c_{21}=$4.9 are the cutoff values of $R_{21}$ and $R_{22}$ under the rule 2, with $\underline{\delta}_{11}=$0.4, $\overline{\delta}_{11}=$0.5, $\underline{\delta}_{12}=$0.1, $\overline{\delta}_{12}=$0.2, $\underline{\delta}_{21}=$0.4, $\overline{\delta}_{21}=$0.5, $\underline{\delta}_{22}=$0.1, and $\overline{\delta}_{22}=$0.2.

\vspace{-.5cm}
\begin{table}[htbp]
\begin{center}
\caption{Selection of a neighborhood for analysis}
\label{table:table2}
\scalebox{0.7}{
\begin{threeparttable}[t]
\centering
\begin{tabular}{@{\extracolsep{5pt}}c c c c c c c@{}}
\hline
\multicolumn{2}{c}{Neighborhood} & \multicolumn{2}{c}{Matched sample size} & $p$-value for the &$p$-value for no effect & Plausibility of \\
\cline{1-2} \cline{3-4}
Rule 1 & Rule 2 & Treated & Control &  cross-match test &on secondary covariates & Assumption 1a\\
\hline
$\text{[3.8, 4.1]} \times \text{[4.3, 4.6]}$ & $\text{[3.8, 4.1]} \times \text{[4.8, 5.1]}$ &  243 & 243 & 1.00 & 0.33 &  \checkmark \\
$\text{[3.7, 4.2]} \times \text{[4.3, 4.6]}$ & $\text{[3.7, 4.2]} \times \text{[4.8, 5.1]}$ &  524 & 524 & 1.00 & 0.87 &   \checkmark \\
$\text{[3.6, 4.3]} \times \text{[4.3, 4.6]}$ & $\text{[3.6, 4.3]} \times \text{[4.8, 5.1]}$ &  841 & 841 & 1.00 & 0.81 &   \checkmark \\
$\text{[3.5, 4.4]} \times \text{[4.3, 4.6]}$ & $\text{[3.5, 4.4]} \times \text{[4.8, 5.1]}$ &  1,141 & 1,141 & 1.00 & 0.16 &   \checkmark \\
$\text{[3.4, 4.5]} \times \text{[4.3, 4.6]}$ & $\text{[3.4, 4.5]} \times \text{[4.8, 5.1]}$ &  1,359 & 1,359 & 1.00 & 0.03 &   \xmark \\
$\text{[3.3, 4.6]} \times \text{[4.3, 4.6]}$ & $\text{[3.3, 4.6]} \times \text{[4.8, 5.1]}$ &  1,578 & 1,578 & 1.00 & 0.01 &   \xmark \\
\hline
\end{tabular}
Notes: the third to last column reports the $p$-value for the Cross-match Test for comparing two
multivariate distributions, while the the second to last column reports the minimum $p$-value for the permutational $t$-test for matched pairs of differences on secondary covariates.
\end{threeparttable}
}
\end{center}
\end{table}

\vspace{-.5cm}
Because it does not require outcome information, this procedure for selecting the neighborhood is part of the design (as opposed to the analysis) stage of the study \citep{rubin2008objective}.
As such, it can help investigators preserve the objectivity of the study and maintain the validity of its statistical tests.
In this procedure, the choice of the covariates in $\boldsymbol{X}$ and $\boldsymbol{X}^{\textrm{test}}$ requires careful consideration of the treatment assignment mechanism in the problem at hand.
In our study, the assignment mechanism can vary across schools and grades, hence we matched exactly for both of these covariates in $\boldsymbol{X}$.
Conditional on these covariates, around a neighborhood of the cutoffs, the student's educational readiness can play an important role in the assignment, hence we further adjusted for all the available standardized test scores.\footnote{As in any observational study based on the assumption of strong ignorability, a possible criticism of our implementation is that we are failing to adjust for a relevant covariate (for example, beyond the household income and education of both parents, their employment status). In this case, it is possible that our estimates are biased. This is something we study through a sensitivity analysis to unmeasured bias.}

Given all these covariates in $\boldsymbol{X}$, we have applied the ideas in \citeauthor{imbens2015causal} (\citeyear{imbens2015causal}, Chapter 21) for assessing unconfoundedness.
Ideally, $\boldsymbol{X}^{\textrm{test}}$ will include variables known not to be affected by the treatment given the covariates in $\boldsymbol{X}$.
These can be proxy or pseudo-outcomes, such as lagged outcomes measured before treatment.
In our data set, we do not have lagged measures of all the study outcomes, but a proxy for them are the measures of social capital corresponding to the mother and father's education, and the household per capita income.
Here, the decision of which variables to include in $\boldsymbol{X}^{\textrm{test}}$ relies on substantive knowledge about the research question; however, one could consider more data-driven approaches that learn the covariates in $\boldsymbol{X}$ and $\boldsymbol{X}^{\textrm{test}}$ from the data itself.
\textcolor{black}{These approaches, nonetheless, should ponder the importance of separating the design and analysis stages of the study and preserving the validity of the statistical tests if they use the outcomes.}

%%%%%%%%%%%%%%%%%%%%%%%%%%%%%%%%%%%%%%%%%%%
%%%%%%%%%%%%%%%%%%%%%%%%%%%%%%%%%%%%%%%%%%%
\subsection{Three analyses}

Having found the neighborhood, we proceed to estimate the effect of grade retention on subsequent school grades, repeating another grade, dropping out of school, or committing a juvenile crime.
%Within this neighborhood, we matched 1,141 students who repeated the academic year in 2007 (treated students) to 1,141 students who advanced to the next grade (control students).
We consider three different analyses: one exploratory, where we visualize the impact of grade retention on subsequent grades across time; one based on randomization inference, where we perform a sensitivity analysis on an equivalence test; and one that combines approximately balancing weights and regularized regression.
These analyses have contrasting strengths and can be complementary in practice.

%%%%%%%%%%%%%%%%%%%%%%%%%%%%%%%%%%%%%%%%%%%
%%%%%%%%%%%%%%%%%%%%%%%%%%%%%%%%%%%%%%%%%%%
\subsubsection{Visualizing patterns of effects}
\label{sec_visualizing}

\begin{figure}[h!]
\caption{Average school grades of the matched students in the selected neighborhood.}
\vspace{2mm}
\begin{center}
\includegraphics[width=3.85in]{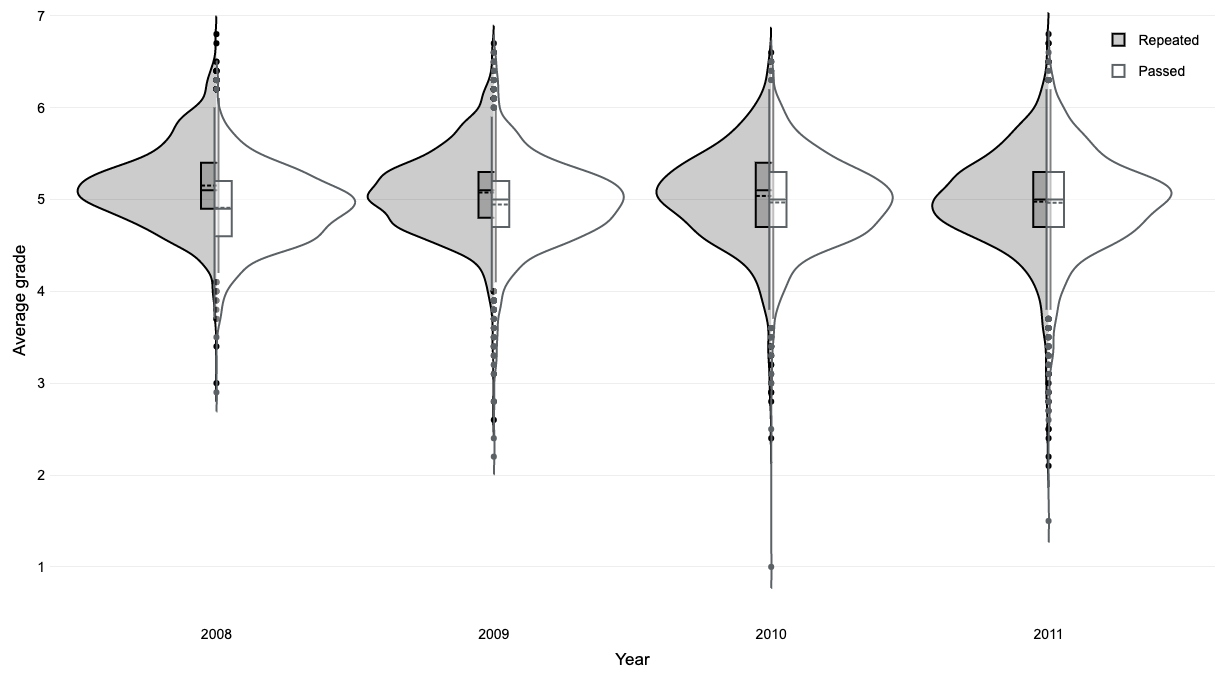}
\end{center}
\label{fig_densities}
\end{figure}

In Figure \ref{fig_densities}, we plot the average school grades of the matched students within the selected neighborhood in years 2008-2011; that is, one, two, three, and four years after repeating or passing in 2007.
In blue, we display the boxplots and densities of the average grades of the students who repeated, and in gray, of the students who passed.
In 2008 (that is, in the year immediately after repeating or passing), the students who repeated had higher average grades than the matched students who passed.
The difference is approximately 0.2 points (see Section \ref{sec_estimating2} for point estimates and confidence intervals).
However, this difference is progressively reduced in the three following years, declining to an average difference of 0.05 points in 2011.
Interestingly, in 2011 the difference of the modes of the distributions of grades appears to be reverted, suggesting that after four years following repeating or passing, there are more students with lower grades after repeating than after passing, nonetheless the opposite happens in 2008.
Between 2008 and 2011 the dispersion of the matched pair differences in outcomes increases from 0.56 to 0.75, suggesting that heterogeneity in treatment effects is increasing with time.\footnote{See Appendix B for a similar visualization of average school grades for students from the same grade, for the first time they were taught the material after grade retention.}

%\vspace{-.5cm}
%\vspace{-.25cm}

%%%%%%%%%%%%%%%%%%%%%%%%%%%%%%%%%%%%%%%%%%%
%%%%%%%%%%%%%%%%%%%%%%%%%%%%%%%%%%%%%%%%%%%
%\vspace{-.75cm}
\subsubsection{Estimating effects using randomization tests}
\label{sec_estimating1}

Here we estimate the effects (risk differences) of grade retention on repeating another grade, dropping out of school, and committing a juvenile crime.
We use the methods in \cite{rosenbaum2002attributing} and \cite{zubizarreta2013stronger}.
%See Appendix B for details.

\begin{figure}[h!]
\caption{Estimates for the Neighborhood Average Treatment Effect}
\vspace{2mm}
\begin{center}
\includegraphics[width=3.9in]{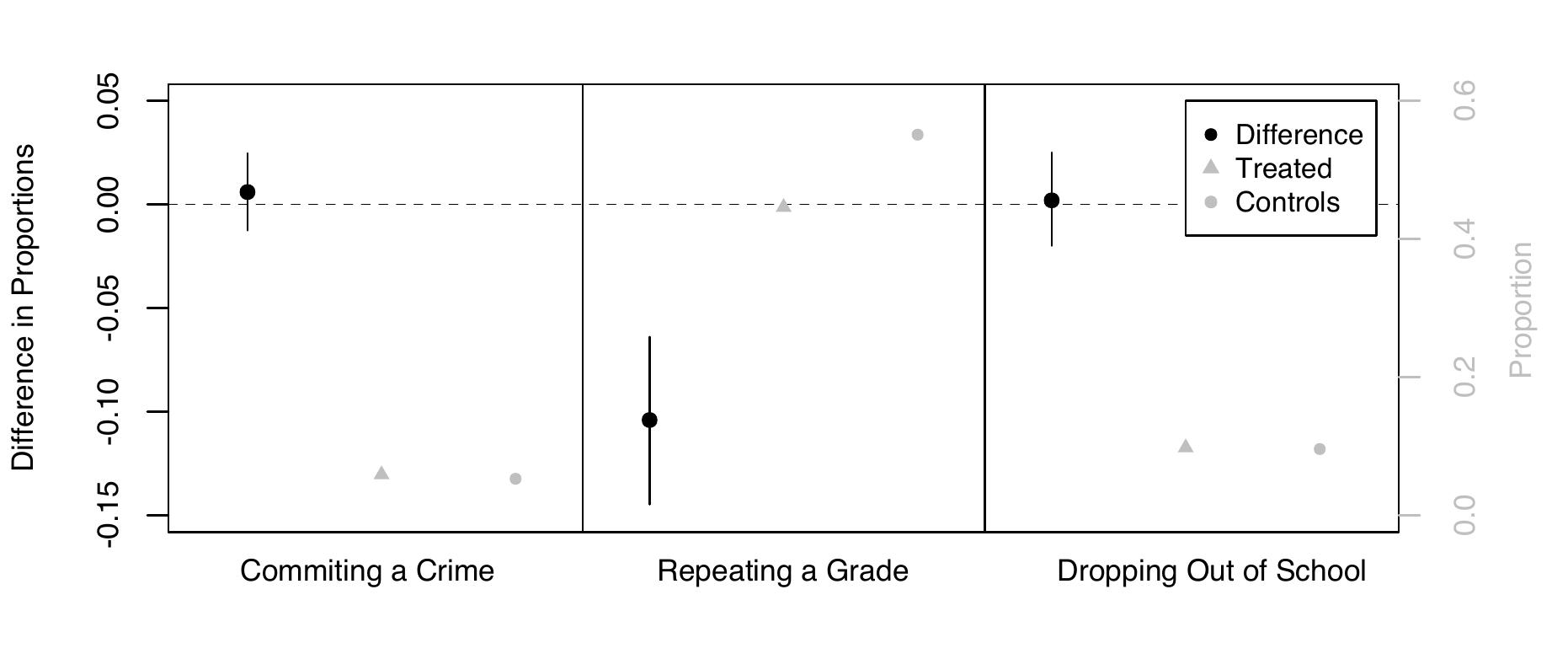}
\end{center}
\label{fig_estimates}
\end{figure}

We test Fisher's null hypothesis of no treatment effects using McNemar's test statistic.
In our example, this test statistic corresponds to the number of discordant pairs where the student that was retained subsequently repeated another grade, dropped out of school, or committed a juvenile crime.
%Table \ref{table:table3} presents the point estimates and $p$-values of the NATE on these three outcome variables.
\textcolor{black}{Figure \ref{fig_estimates} presents the point estimates and confidence intervals of the NATE on these three outcome variables.}
The point estimate of the effect of grade retention on juvenile crime is  $\widehat{\tau}_{\rm{NATE}}^{\rm{crime}}$ = 0.006  \textcolor{black}{($p$-value = 0.229)}.
In fact, among the 1,141 matched pairs, there are 117 discordant pairs in which only one student committed a crime.
Of these, there are 62 pairs in which the student that repeated committed a crime (see Appendix C for details).
Thus, in the absence of hidden bias, there is no evidence that grade retention causes juvenile crime.
Similarly, the estimated effect of grade retention on dropping out of school is also very small ($\widehat{\tau}_{\rm{NATE}}^{\rm{drop}}$ = 0.002) and not statistically significant ($p$-value = 0.379).
However, the estimated effect of grade retention on repeating another grade is $\widehat{\tau}_{\rm{NATE}}^{\rm{ret}}$ = -0.104.
Here, there are 563 discordant pairs, and out of these, there are 222 pairs in which the student who was retained in 2007 did not repeat another grade in the future, yielding a one-sided $p$-value smaller than 0.001 (see Appendix C).
Thus, in the absence of hidden bias, there is significant evidence that current grade retention causes a reduction in future grade retention.
In Appendix H, we analyze the stability of these results to varying the neighborhood size\textcolor{black}{, finding that they are  generally robust to modifying the neighborhood for the lowest and second lowest grades, but somewhat sensitive to modifying the neighborhood for the average grade,\footnote{For these values, however, the covariate balance conditions are not satisfied (hence, these neighborhoods would not be eligible for analysis by the discussed neighborhood selection procedure).} emphasizing the importance of the selection of the neighborhood.}
In Appendix F, we provide guidance for generalizing these results.
%\vspace{-.25cm}

%\vspace{-.5cm}
We also study how these effects vary by gender.
This is straightforward because we matched exactly by that covariate.
The results are reported in Tables 4 and 5 in Appendix C, and suggest that the effect of grade retention is not modified by gender.
Under Assumptions \ref{as_local_strong_ignorability}\textcolor{black}{a} and \ref{as_local_strong_ignorability}\textcolor{black}{b} it is possible to study treatment effect heterogeneity in the discontinuity neighborhood using other regression approaches rather than matching.

In the absence of hidden bias, we have found evidence that grade retention does not cause dropping out of school or committing a juvenile crime.
However, bias from a hidden covariate can give the impression that a treatment effect does not exist when in fact there is one.
How much bias from a hidden covariate would need to be present to mask an actual treatment effect?
We address this question with a sensitivity analysis on a near equivalence test (\citealp{rosenbaum2009sensitivity}, \citealp{zubizarreta2013stronger}).
Our results reveal that two students matched on their observed covariates could differ in their odds of repeating the grade in 2007 by almost 12\% and 46\% before masking small and moderate effects previously documented in the literature on juvenile crime.
Analogously, two students matched for their covariates could differ in their odds of repeating by 16\% and 33\% before masking small and moderate effects on dropping out of school.
See Appendices D and E for details.

%%%%%%%%%%%%%%%%%%%%%%%%%%%%%%%%%%%%%%%%%%%
%%%%%%%%%%%%%%%%%%%%%%%%%%%%%%%%%%%%%%%%%%%
\subsubsection{Estimating effects using balancing weights and regularized regression}
\label{sec_estimating2}

We use the regression assisted weighting estimator (\ref{eq_raw}) to estimate the effect of grade retention on average school grades in subsequent school grades.
More specifically, we use the method by \cite{athey2018approximate} which combines approximately balancing weights and \textcolor{black}{regularized linear regression}.
We find that one year after retention, the average school grades of students who repeated is 0.19 points higher than the one of students that passed (95\% confidence interval, [0.16, 0.21]; the standard deviation of average school grades this year was 0.46 points).
Four years after retention, this effect is considerably reduced, and students who repeated have on average 0.05 points higher than students who passed (95\% confidence interval, [0.01, 0.08]; the standard deviation of average school grades in that year was 0.63 points).
%\textcolor[rgb]{1.00,0.00,0.00}{Para interpretar los resultados, argumentar\'ia algo similar a los expuesto en el an\'alisis gr\'afico.}

In summary, our results suggest that grade retention has a positive effect in the short run on future school grades, but that this effect dissipates over time --- all this, among students who are comparable in terms of observed covariates \textcolor{black}{and} that barely pass or repeat the academic year.
\textcolor{black}{A number of mechanisms can explain this result.}
\textcolor{black}{One of them} is that retained students know the material better and are more mature.
This is consistent with our other result that students who barely pass tend to repeat more in the future.
\textcolor{black}{It is also possible that school grades measure the students' true ability with error, so that in the neighborhood below the cutoff, we have more students who suffered a negative transitory shock, while above the cutoff we have more students who suffered a positive transitory shock.
In that case, it is possible that students that barely repeat are more likely to improve next year, and vice-versa, even if retention has no effect.
Although it merits further investigation, this explanation may be limited as we adjust for educational and socioeconomic variables in the four years preceding the intervention of grade retention.
This and other possible causal mechanisms that may explain our findings deserve closer examination.}
Finally, we estimate an almost null effect of grade retention on juvenile crime, with these results being insensitive to small and moderate hidden biases.

%% file: dd89_sec7_01.tex
The proposed framework can also accommodate intermediate outcomes (i.e., principal stratification analyses).
For example, it can be extended to instrumental variable (IV) analyses with noncompliance \citep{angrist1996identification}, yet within a neighborhood of the cutoffs of the running variables.
As mentioned, in discontinuity designs the IV approach to noncompliance is known as the \emph{fuzzy} design, whereas the case with full compliance is known as the \emph{sharp} design \citep{hahn2001identification}.
In a complex fuzzy discontinuity design (that is, when the treatment assignment is determined by multiple rules that possibly lead to the same treatment, but there is noncompliance), we can nonparametrically identify the average treatment effect for the compliers in the discontinuity neighborhood under two assumptions beyond assumptions \ref{as_local_strong_ignorability}\textcolor{black}{a} and \ref{as_local_strong_ignorability}\textcolor{black}{b}.
Specifically, under the assumptions of local strong ignorability of treatment assignment (\ref{as_local_strong_ignorability}\textcolor{black}{a} and \ref{as_local_strong_ignorability}\textcolor{black}{b}), local monotonicity (i.e., that the treatment assignment has a non-negative effect on the treatment exposure for all the units in a neighborhood of the cutoffs), and the local exclusion restriction (i.e., that the treatment assignment affects the outcome only through the treatment exposure for all the units the neighborhood), the average treatment effect for the compliers in the discontinuity neighborhood can be identify following the same arguments as those in the IV approach to noncompliance in randomized experiments (\citealp{imbens2015causal}, chapters 23 and 24), but restricting the analysis to a neighborhood of the cutoffs.

%% file: dd89_sec8_01.tex
Since their introduction in 1960 by \citeauthor{thistlethwaite1960regression}, regression discontinuity designs have proven to be a powerful method for drawing causal inferences in observational studies.
However, they have often been confined to settings where treatment assignment is determined by simple rules.
Although regression discontinuity designs under the continuity-based framework have been extended to separately incorporate multiple running variables or multiple cutoffs, to our knowledge they do not comprehend more complex treatment rules, such as those found in our case study.
In this paper we have conceptualized a complex discontinuity design as a local randomized experiment conditional on covariates; or more specifically, as an observational study with strongly ignorable treatment assignment given covariates in a neighborhood of the cutoffs.
In our case study, we found that grade retention in Chile has a negative impact on future grade retention, but is not associated with dropping out of school or committing a juvenile crime.
As discussed, this framework can facilitate the generalization and transportation of causal inferences in discontinuity designs.
Under this framework, it is straightforward to take advantage of methods not normally used in traditional regression discontinuity designs, such as simple graphical displays of outcomes as in clinical trials, and potentially statistical machine learning methods to estimate heterogeneous effects.
Here, we have used matching to adjust for covariates and select the neighborhood, but under the assumptions of local strong ignorability other methods can be used.
Future work can study the properties of regression assisted estimators in discontinuity designs and extend this framework to principal stratification analyses (see \citealp{li2015evaluating}).
In observational studies, discontinuities in treatment assignment rules offer a keyhole through which to see causality.
Motivated by the problem of estimating the impact of grade retention on later life outcomes, in this paper we have proposed a framework and methods to leverage them with complex treatment rules.

%% file: dd89arxiv_supplementary.tex
The following materials are organized as follows.
Appendix A describes covariate balance before and after matching.
Appendix B displays the outcomes after matching for students in the same grade.
Appendix C includes additional explanations and results on the outcome analyses.
Appendices D and E expand on the sensitivity analyses.
Appendix F discusses generalization.
Appendix G explains how to implement the design-based approach for selecting a neighborhood, presents a separate semi-design-based approach, and provides some practical considerations. 
Appendix H evaluates the stability of the results to varying the neighborhood size.
Appendix I analyzes covariate balance under the continuity-based framework.
%\end{abstract}
%
%%\vspace*{.3in}
%
%\begin{center}
%\noindent Keywords:
%%\small
%{Causal Inference; Observational Studies; Regression Discontinuity Design}
%%\normalsize
%\end{center}
%\clearpage
%\doublespacing

%%%%%%%%%%%%%%%%%%%%%%%%%%%%%%%%%%%%%%%%%%%
%%%%%%%%%%%%%%%%%%%%%%%%%%%%%%%%%%%%%%%%%%%
%%%%%%%%%%%%%%%%%%%%%%%%%%%%%%%%%%%%%%%%%%%
\pagebreak
%\setcounter{page}{1}
%\section*{Online Supplementary Materials}
\section*{Appendix A: Covariate balance}

\begin{figure}[htbp]
\begin{center}
\caption{Standardized differences in means before and after matching inside the neighborhood $[3.5, 4.4]\times [4.3, 4.6]$ for rule 1 and $[3.5, 4.4] \times [4.8, 5.1]$ for rule 2.}\label{loveplot}
\includegraphics[scale=0.5]{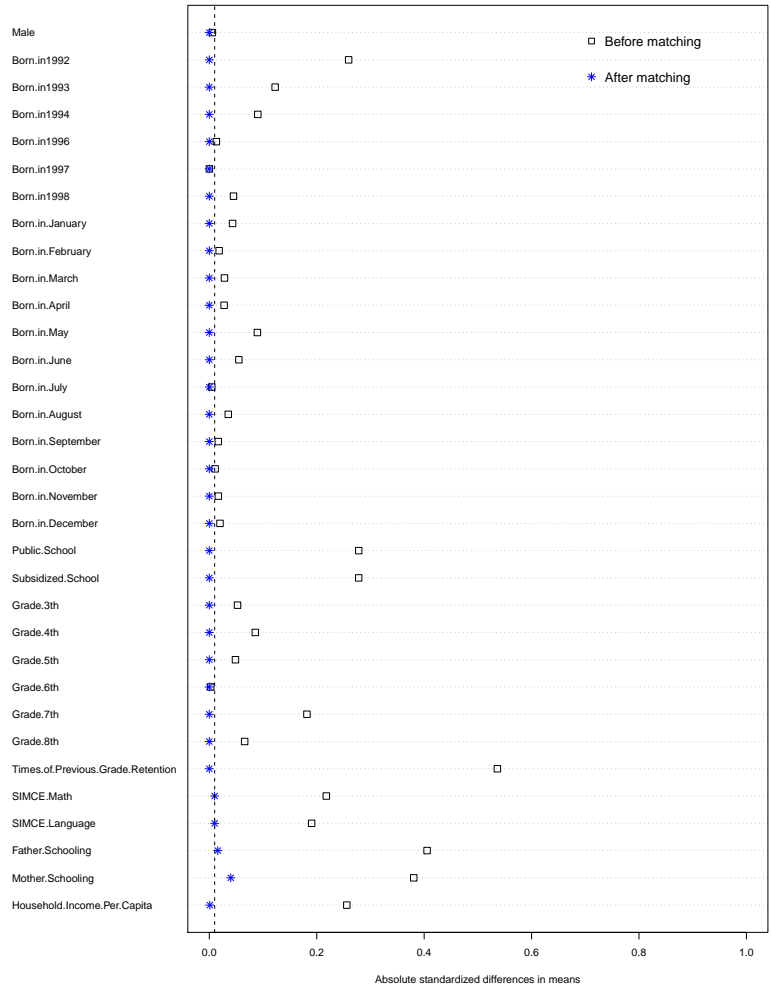}
\floatfoot{Notes: inside the neighborhood, we match exactly on gender, year and month of birth, school and grade attended in 2007, and times of past grade retention.  In addition, we match with mean balance on the SIMCE scores in language and mathematics.  As a result, the mother and father's schooling and the household income are also balanced.}
\end{center}
\end{figure}

%%%%%%%%%%%%%%%%%%%%%%%%%%%%%%%%%%%%%%%%%%%
%%%%%%%%%%%%%%%%%%%%%%%%%%%%%%%%%%%%%%%%%%%
%%%%%%%%%%%%%%%%%%%%%%%%%%%%%%%%%%%%%%%%%%%
\pagebreak
\section*{Appendix B: Patterns of effects}

\begin{figure}[h!]
\caption{Average school grades of the matched students in the selected neighborhood. Here, we compare grades for the matched treated and control students from the same grade, for the first time they were taught that material.}
\vspace{2mm}
\begin{center}
\includegraphics[width=3.85in]{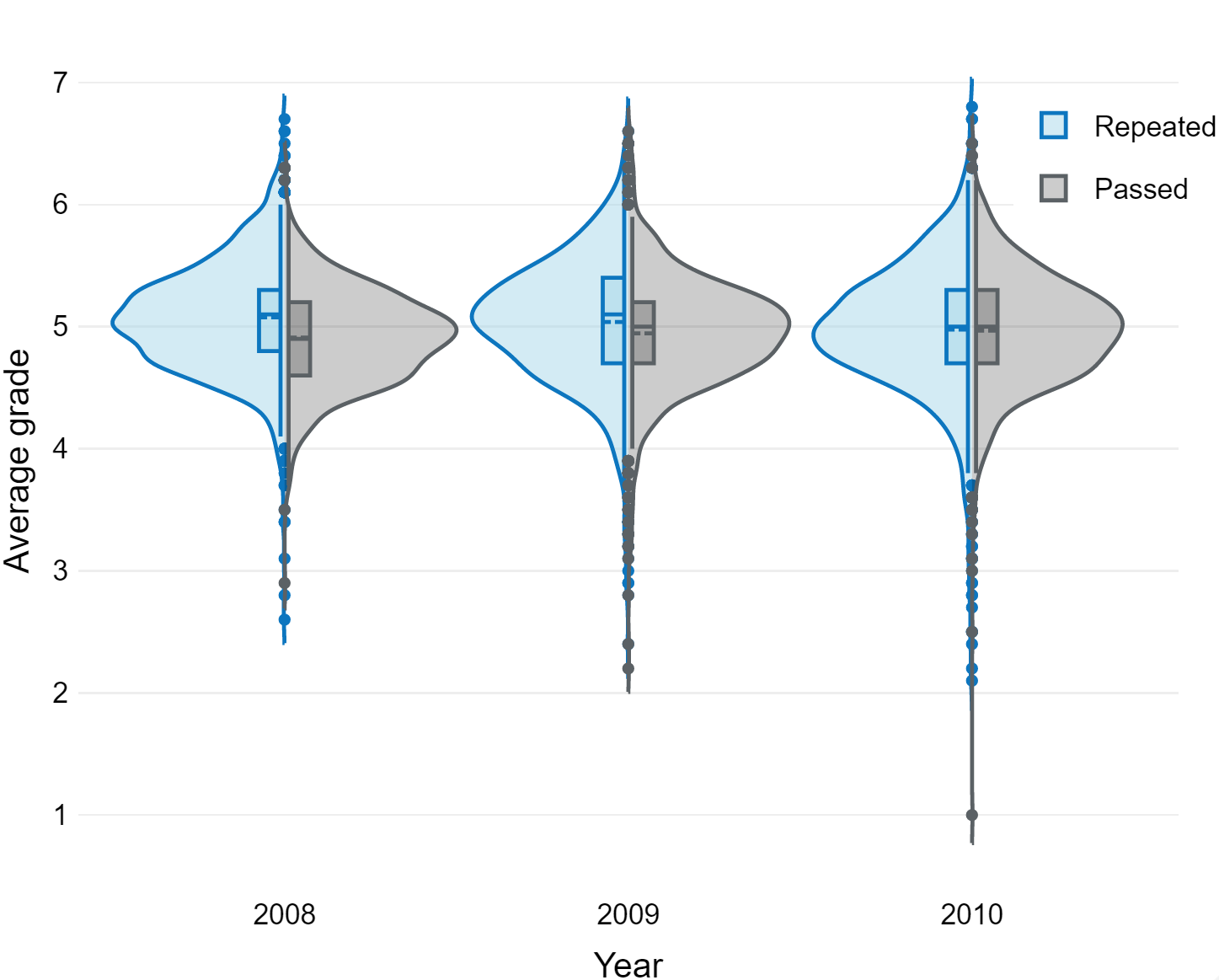}
\end{center}
\label{fig_densities}
\end{figure}

%%%%%%%%%%%%%%%%%%%%%%%%%%%%%%%%%%%%%%%%%%%
%%%%%%%%%%%%%%%%%%%%%%%%%%%%%%%%%%%%%%%%%%%
%%%%%%%%%%%%%%%%%%%%%%%%%%%%%%%%%%%%%%%%%%%
\pagebreak
\section*{Appendix C: Outcome analyses}

We have $I=1,141$ matched pairs.
In each pair $i \in \{1,2,\ldots,I\}$, one student $j \in \{1,2\}$ repeated the grade and the other passed.
Following \cite{rosenbaum2002observational}, let $Z_{ij}=1$ if student $j$ in matched pair $i$ repeats and $Z_{ij}=0$ otherwise, so $Z_{i1} + Z_{i2} = 1$ for all $i \in \{1,2,\ldots,I\}$.
Student $j$ in pair $i$ exhibits potential outcome $Y_{ij}(1)$ if $Z_{ij}=1$, and potential outcome $Y_{ij}(0)$ if $Z_{ij}=0$.
In the set $\mathcal{Z}$, we collect the $2^I$ possible treatment assignments, $\boldsymbol{Z} =(Z_{11},Z_{12},\ldots, Z_{I1},Z_{I2})^\top$.
Under paired randomization, $\mathrm{Pr}\{ Z_{ij}=1|Y_{ij}(1),Y_{ij}(0), u_{ij},\boldsymbol{X}_{ij}, N_{ij}=1, \mathcal{Z} \}=1/2$, where $u_{ij}$ is an unobserved covariate.
Let $\boldsymbol{Y}=(Y_{11},Y_{12},\ldots, Y_{I1}, Y_{I2})^\top$ denote the vector of the observed outcomes for the $2I$ students and let $\boldsymbol{Y}(0)=(Y_{11}(0),Y_{12}(0),\ldots, Y_{I1}(0), Y_{I2}(0))^\top$ stand for the vector of potential outcomes under control for the $2I$ students.
If $t(\boldsymbol{Z},\boldsymbol{Y})$ is a test statistic, then in a paired randomized experiment under Fisher's null hypothesis of no treatment effect, the distribution of $t(\boldsymbol{Z},\boldsymbol{Y})$ is the permutation distribution
\begin{equation}\label{prandomization}
\mathrm{Pr}(t(\boldsymbol{Z},\boldsymbol{Y})> k |Y_{ij}(1),Y_{ij}(0), u_{ij},\boldsymbol{X}_{ij}, N_{ij}=1, \mathcal{Z})=\frac{|\{\boldsymbol{Z} \in \mathcal{Z}: t(\boldsymbol{Z},\boldsymbol{Y}(0))>k\}|}{2^I}.
\end{equation}

In our running example, the three outcomes of interest are binary.
The first outcome takes the value 1 if the student committed a crime after 2007 and 0 otherwise; the second takes the value 1 if the student repeated a grade after 2007 and 0 otherwise; and the third takes the value 1 if the student dropped out of school after 2007 and 0 otherwise.
To test $H_0$, we use McNemar's test statistic: $t(\boldsymbol{Z},\boldsymbol{Y})=\sum_{i=1}^{I}\sum_{j=1}^{2}Z_{ij}Y_{ij}$, i.e., the number of responses equal to 1 among treated students.
The results are as follows.

\begin{table}[htbp]
\begin{center}
\caption{Committing a juvenile crime in matched pairs. The table counts pairs, not students.}
\label{table:pairscrime}
\scalebox{0.8}{
  \centering
 \begin{tabular}{l c  c}
 \hline
  &\multicolumn{2}{c}{Passed}  \\
 \cline{2-3}
Repeated & Juvenile crime = 0 & Juvenile crime = 1 \\
 \hline
 Juvenile crime = 0  &  1018 & 55\\
 Juvenile crime = 1  &  62 & 6 \\
 \hline
\end{tabular}
}
\end{center}
\end{table}

\begin{table}[htbp]
\begin{center}
\caption{Dropping out of school in matched pairs. The table counts pairs, not students.}
\label{table:pairsdrop}
\scalebox{0.8}{
  \centering
 \begin{tabular}{l c  c}
 \hline
  &\multicolumn{2}{c}{Passed}  \\
 \cline{2-3}
Repeated & Dropped out = 0 & Dropped out = 1 \\
 \hline
 Dropped out = 0  &  945 & 84\\
 Dropped out = 1  &  87 & 25 \\
 \hline
\end{tabular}
}
\end{center}
\end{table}

\begin{table}[htbp]
\begin{center}
\caption{Repeating another grade in matched pairs. The table counts pairs, not students.}
\label{table:pairsret}
\scalebox{0.8}{
  \centering
 \begin{tabular}{l c  c}
 \hline
  &\multicolumn{2}{c}{Passed}  \\
 \cline{2-3}
Repeated & Future retention = 0 & Future retention = 1 \\
 \hline
 Future retention = 0  &  290 & 341\\
 Future retention = 1  &  222 & 288 \\
 \hline
\end{tabular}
}
\end{center}
\end{table}

\begin{table}[htbp]
\begin{center}
\caption{Estimates for the Neighborhood Average Treatment Effect, Boys (689 pairs)}
\label{table:table3}
\scalebox{0.8}{
\begin{threeparttable}[t]
\centering
\begin{tabular}{l p{2.2cm}  p{2.2cm}  p{2.2cm} c }
\hline
\multirow{2}{*}{Outcome variable} &\multicolumn{2}{c}{Matched sample mean} & \multicolumn{1}{c}{\multirow{2}{*}{$\widehat{\tau}_{\rm{NATE}}$}}  & $H_0: \tau_{\rm{NATE}}=0$ \\
\cline{2-3}
& \multicolumn{1}{c}{Treated} & \multicolumn{1}{c}{Control} &  & One-sided $p$-value \\
\hline
Committing a crime  &  \multicolumn{1}{c}{0.082} & \multicolumn{1}{c}{0.075} & \multicolumn{1}{c}{\phantom{-}0.007}  & \phantom{<}0.307\\
Repeating another grade  &  \multicolumn{1}{c}{0.474} & \multicolumn{1}{c}{0.577} & \multicolumn{1}{c}{-0.103}  & <0.001\\
 Dropping out of school  &  \multicolumn{1}{c}{0.103} & \multicolumn{1}{c}{0.108} & \multicolumn{1}{c}{-0.005} & \phantom{<}0.351\\
\hline
\end{tabular}
\end{threeparttable}
}
\end{center}
\end{table}

\begin{table}[htbp]
\begin{center}
\caption{Estimates for the Neighborhood Average Treatment Effect, Girls (452 pairs)}
\label{table:table3}
\scalebox{0.8}{
\begin{threeparttable}[t]
\centering
\begin{tabular}{l p{2.2cm}  p{2.2cm}  p{2.2cm} c }
\hline
\multirow{2}{*}{Outcome variable} &\multicolumn{2}{c}{Matched sample mean} & \multicolumn{1}{c}{\multirow{2}{*}{$\widehat{\tau}_{\rm{NATE}}$}}  & $H_0: \tau_{\rm{NATE}}=0$ \\
\cline{2-3}
& \multicolumn{1}{c}{Treated} & \multicolumn{1}{c}{Control} &  & One-sided $p$-value \\
\hline
Committing a crime  &  \multicolumn{1}{c}{0.024} & \multicolumn{1}{c}{0.020} & \multicolumn{1}{c}{\phantom{-}0.004}  & \phantom{<}0.381\\
Repeating another grade  &  \multicolumn{1}{c}{0.404} & \multicolumn{1}{c}{0.511} & \multicolumn{1}{c}{-0.106}  &  \phantom{<}0.001\\
 Dropping out of school  &  \multicolumn{1}{c}{0.901} & \multicolumn{1}{c}{0.075} & \multicolumn{1}{c}{\phantom{-}0.015} & \phantom{<}0.185\\
\hline
\end{tabular}
\end{threeparttable}
}
\end{center}
\end{table}

%%%%%%%%%%%%%%%%%%%%%%%%%%%%%%%%%%%%%%%%%%%
%%%%%%%%%%%%%%%%%%%%%%%%%%%%%%%%%%%%%%%%%%%
%%%%%%%%%%%%%%%%%%%%%%%%%%%%%%%%%%%%%%%%%%%
\pagebreak
\section*{Appendix D: Rosenbaum bounds to assess sensitivity to hidden bias}

In the absence of hidden bias, there is strong evidence in our running example that current grade retention causes a reduction in future grade retention.
How much hidden bias would need to be present to explain away this result?
To answer this question, we implement the sensitivity analysis described in \citeauthor{rosenbaum2002observational} (2002b, Chapter 4).

Let $\pi_{ij}$ denote the probability that student $j$ in pair $i$ repeats the grade (receives treatment).
For each each pair $i$, two students match on their observed covariates, $\boldsymbol{X}_{ij} =\boldsymbol{X}_{ij^{'}}$, but may differ on an unobserved covariate $u_{ij} \neq u_{ij^{'}}$, such that $\pi_{ij} \neq\pi_{ij^{'}}$.
Suppose that the odds of repeating the grade differ at most by a factor $\Gamma\geq 1$
\begin{equation*}\label{sensibility}
\frac{1}{\Gamma}\leq \frac{\pi_{ij}(1-\pi_{ij^{'}})}{\pi_{ij^{'}}(1-\pi_{ij})}\leq \Gamma
\end{equation*}
for each pair $i$.
If $\Gamma=1$, then there is no hidden bias and the randomization distribution with $\pi_{ij}=\pi_{ij^{'}}=1/2$ for McNemar's test statistic is valid.
If $\Gamma > 1$, then there is hidden bias and there is a range of possible inferences for $\pi_{ij} \neq\pi_{ij^{'}}$.
These inferences are bounded by $\Gamma$ and $1/\Gamma$.
For these two values, we obtain two extreme-case $p$-values.
We look for the largest value of $\Gamma$ such that we reject the null hypothesis of no treatment effect.

In our running example, we are able to reject the null hypothesis that current grade retention does not causes future grade retention for $\Gamma=1.34$ (the upper bound of the $p$-value is 0.049) but not for $\Gamma>1.34$ (the upper bound of the $p$-value is greater than 0.05).
In other words, two students matched for their observed covariates could differ in their odds of grade retention by 34\% without materially altering the conclusions about the effect of current retention on future grade retention.

%%%%%%%%%%%%%%%%%%%%%%%%%%%%%%%%%%%%%%%%%%%
%%%%%%%%%%%%%%%%%%%%%%%%%%%%%%%%%%%%%%%%%%%
%%%%%%%%%%%%%%%%%%%%%%%%%%%%%%%%%%%%%%%%%%%
\pagebreak
\section*{Appendix E: Sensitivity analysis on a near equivalence test}

As discussed in Section 6.2.2, in the absence of hidden bias, we have found evidence that grade retention does not cause dropping out of school or committing a juvenile crime.
However, bias from a hidden covariate can give the impression that a treatment effect does not exist when in fact there is one.
How much bias from a hidden covariate would need to be present to mask an actual treatment effect?
We answer this question by conducting a sensitivity analysis on a near-equivalence test.
For details, see Rosenbaum and Silber (2009) and Zubizarreta et al. (2013).
%\cite{rosenbaum2009sensitivity} and \cite{zubizarreta2013stronger}.

Using the parameter $\Gamma$, we will test the null hypothesis that the effect of treatment is larger than a given effect size against the alternative hypothesis that it is lower.
We take the effect size from previous studies in the literature.
Our results will be insensitive to hidden bias if large values of $\Gamma$ are required to mask a given effect size.
Before presenting the results of this sensitivity analysis, it is necessary to introduce the concept of an attributable effect \citep{rosenbaum2002attributing}.

For each of the two outcomes, committing a juvenile crime or dropping out of school, let $\bm{\delta}$ denote the $2I$-dimensional vector of treatment effects, with $\delta_{ij}=Y_{ij}(1)-Y_{ij}(0)$.
Consider the hypothesis $H_{\bm{\delta_0}}: \bm{\delta}=\bm{\delta_0}$, where $\bm{\delta_0}$ is a $2I$-dimensional vector with values $\delta_{0ij} \in \{-1,0,1\}$.
Clearly, not all hypotheses of this form are consistent with the data at hand, since  $Y_{ij}-Z_{ij}\delta_{0ij}=Y_{ij}(0)$ and $Y_{ij}+(1-Z_{ij})\delta_{0ij}=Y_{ij}(1)$ must both be in $\{0,1\}$.
%Although testing $H_{\bm{\delta_0}}: \bm{\delta}=\bm{\delta_0}$ is not difficult, a complication appears in practice when testing many values of $\bm{\delta_0}$. To overcome this practical issue,
\cite{rosenbaum2002attributing} summarizes hypotheses of the form $H_{\bm{\delta_0}}: \bm{\delta}=\bm{\delta_0}$ by introducing the attributable effect, which is defined as $\Delta=\sum_{i=1}^{I}\sum_{j=1}^{2}Z_{ij}\delta_{ij}$.
In other words, the attributable effect is the number of treated students who experienced events caused by the treatment.
%This quantity is a random variable because it depends on $\boldsymbol{Z}$, but it is unobserved because $\bm{\delta}$ is unknown.

We assume that $Y_{ij}(1) \geq Y_{ij}(0)$ for each student $j$ in pair $i$; that is, that the exposure to treatment may cause the outcome in a student who would not otherwise experience it, but the treatment does not prevent the outcome in a student who would otherwise experience it.
We consider the hypothesis $H_{\bm{\delta_0}}: \bm{\delta}=\bm{\delta_0}$, where $\delta_{0ij} \in \{0,1\}$.
%This is not a restricted conditions because
We can repeat this exercise reversing the roles of treatment and control.
We again use McNemar's test statistic.
Under $H_{\bm{\delta_0}}: \bm{\delta}=\bm{\delta_0}$, $\Delta$ may be calculated using $\bm{\delta_0}$, $\Delta_0=\sum_{i=1}^{I}\sum_{j=1}^{2}Z_{ij}\delta_{0ij}$ and $t(\boldsymbol{Z},\boldsymbol{Y}(0))$ can be computed.
A value of $\Delta_0$ is rejected if every hypothesis $H_{\bm{\delta_0}}: \bm{\delta}=\bm{\delta_0}$ with $\delta_{0ij} \in \{0,1\}$ that gives rise to this value of $\Delta_0$ is rejected.
Given $\Delta_0$, among all the hypotheses $H_{\bm{\delta_0}}: \bm{\delta}=\bm{\delta_0}$ with $\delta_{0ij} \in \{0,1\}$ that give rise to this value of $\Delta_0$, there is one that is the most difficult to reject,
such that if it is rejected then the associated value of $\Delta_0$ is rejected.
\cite{rosenbaum2001effects} proves that the hypothesis that is the most difficult to reject has $\sum_{j=1}^{2}Z_{ij}\delta_{ij}=1$ for as many pairs with $Y_{i1}+Y_{i2}=2$ as possible.

When conducting the test of equivalence on the matched pairs, we use two values for the attributable effect on juvenile crime ($\Delta^{crime}_0 \in \{30,40\}$) and also two values for the attributable effect on dropping out of school ($\Delta^{drop}_0 \in \{40,50\}$). For example, an attributable effect on juvenile crime of 40 is equivalent to saying that 40 juvenile crimes are attributed to grade retention.
The order of magnitude of these attributable effects are taken from previous studies in the literature.
Utilizing a standard fuzzy RDD, Erena et al. (2017)
%\cite{erena2017test}
estimate the impact of grade retention on dropping out of school in the state of Louisiana, finding a point estimate of 4.8\% (equivalent to 55 cases of dropping out of school attributed to grade retention in our context), whereas \cite{grau2018crime} estimate the effect of grade retention on juvenile crime in Chile, obtaining a point estimate of 4.3\% (equivalent to 49 juvenile crimes attributed to grade retention in our application).

We are able to reject the null hypothesis that $H_0: \Delta^{crime}\geq 30$ for $\Gamma=1.12$ with an upper bound of the $p$-value of 0.049 (Table \ref{table:adjparirscrime30} presents the adjusted pairs), whereas we reject the null hypothesis that $H_0: \Delta^{crime}\geq 40$ for $\Gamma=1.46$ with an upper bound of the $p$-value of 0.047 (Table \ref{table:adjparirscrime40} presents the adjusted pairs). These results reveal that two students matched on the basis of their observed covariates could differ in their odds of repeating the grade in 2007 by almost 12\% before masking an attributable effect on juvenile crime of 30, and by almost 46\% before masking an attributable effect of 40. In the same manner, $H_0: \Delta^{drop}\geq 40$ is rejected for $\Gamma=1.16$ with the upper bound of the $p$-value of 0.047 (Table \ref{table:adjparirsdrop40} presents the adjusted pairs), whereas $H_0: \Delta^{drop}\geq 50$ is also rejected for $\Gamma=1.33$ with the upper bound of the $p$-value of 0.045 (Table \ref{table:adjparirsdrop50} presents the adjusted pairs). In words, these results suggest that two students matched for their observed covariates could differ in their odds of repeating the grade in 2007 by almost 16\% before masking an attributable effect of 40 on dropping out of school, and by almost 33\% before masking an attributable effect of 50.

\begin{table}[htbp]
\begin{center}
\caption{Juvenile crime indicators in matched pairs adjusted for the null
hypothesis $H_0: \bm{\delta^{crime}}=\bm{\delta_0^{crime}}$ that attributes $\sum_{i=1}^{I}\sum_{j=1}^{2}Z_{ij}\delta_{ij}^{crime}=30$ juvenile crimes to grade retention. The table counts pairs, not students.}
\label{table:adjparirscrime30}
\scalebox{0.8}{
  \centering
 \begin{tabular}{l c  c}
 \hline
  &\multicolumn{2}{c}{Passed}  \\
 \cline{2-3}
Repeated & Juvenile crime = 0 & Juvenile crime = 1 \\
 \hline
 Juvenile crime = 0  &  1042 & 61\\
 Juvenile crime = 1  &  38 & 0 \\
 \hline
\end{tabular}
}
\end{center}
\end{table}

\begin{table}[htbp]
\begin{center}
\caption{Juvenile crime indicators in matched pairs adjusted for the null
hypothesis $H_0: \bm{\delta^{crime}}=\bm{\delta_0^{crime}}$ that attributes $\sum_{i=1}^{I}\sum_{j=1}^{2}Z_{ij}\delta_{ij}^{crime}=40$ juvenile crimes to grade retention. The table counts pairs, not students.}
\label{table:adjparirscrime40}
\scalebox{0.8}{
  \centering
 \begin{tabular}{l c  c}
 \hline
  &\multicolumn{2}{c}{Passed}  \\
 \cline{2-3}
Repeated & Juvenile crime = 0 & Juvenile crime = 1 \\
 \hline
 Juvenile crime = 0  &  1052 & 61\\
 Juvenile crime = 1  &  28 & 0 \\
 \hline
\end{tabular}
}
\end{center}
\end{table}

\begin{table}[htbp]
\begin{center}
\caption{Dropping out of school indicators in matched pairs adjusted for the null
hypothesis $H_0: \bm{\delta^{drop}}=\bm{\delta_0^{drop}}$ that attributes $\sum_{i=1}^{I}\sum_{j=1}^{2}Z_{ij}\delta_{ij}^{drop}=40$ cases  of dropping out of school to grade retention. The table counts pairs, not students.}
\label{table:adjparirsdrop40}
\scalebox{0.8}{
  \centering
 \begin{tabular}{l c  c}
 \hline
  &\multicolumn{2}{c}{Passed}  \\
 \cline{2-3}
Repeated & Dropped out = 0 & Dropped out = 1 \\
 \hline
 Dropped out = 0  &  960 & 109\\
 Dropped out = 1  &  72 & 0 \\
 \hline
\end{tabular}
}
\end{center}
\end{table}

\begin{table}[t]
\begin{center}
\caption{Dropping out of school indicators in matched pairs adjusted for the null
hypothesis $H_0: \bm{\delta^{drop}}=\bm{\delta_0^{drop}}$ that attributes $\sum_{i=1}^{I}\sum_{j=1}^{2}Z_{ij}\delta_{ij}^{drop}=50$ cases  of dropping out of school to grade retention. The table counts pairs, not students.}
\label{table:adjparirsdrop50}
\scalebox{0.8}{
  \centering
 \begin{tabular}{l c  c}
 \hline
  &\multicolumn{2}{c}{Passed}  \\
 \cline{2-3}
Repeated & Dropped out = 0 & Dropped out = 1 \\
 \hline
 Dropped out = 0  &  970 & 109\\
 Dropped out = 1  &  62 & 0 \\
 \hline
\end{tabular}
}
\end{center}
\end{table}

%%%%%%%%%%%%%%%%%%%%%%%%%%%%%%%%%%%%%%%%%%%
%%%%%%%%%%%%%%%%%%%%%%%%%%%%%%%%%%%%%%%%%%%
%%%%%%%%%%%%%%%%%%%%%%%%%%%%%%%%%%%%%%%%%%%
\clearpage
\pagebreak
\section*{Appendix F: Generalizing the results}

Here we generalize our results to a target population.
Under the Assumptions 1a, 1b, 2a, and 2b, we can estimate the average treatment effect on a target population with running variables in the neighborhood.
We define the target population on the basis of household income.
Among students with values of the running variables in the neighborhood previously determined, the target population consists of students in the poorest 50\% of households.

\begin{table}[h]
\begin{center}
\caption{Description of the target population and the matched sample within the window $[3.5, 4.4]\times [4.3, 4.6]$ for rule 1 and $[3.5, 4.4] \times [4.8, 5.1]$ for rule 2.}
\label{table:balance1}
\scalebox{0.8}{
\begin{threeparttable}[t]
  \centering
 \begin{tabular}{l p{2.2cm} c p{2.2cm} p{2.2cm} }
 \hline
 \multirow{2}{*}{Covariate} &\multicolumn{1}{c}{Full sample mean of the} & & \multicolumn{2}{c}{Matched sample mean} \\
 \cline{4-5}
& \multicolumn{1}{c}{poorest 50\% of households} & &  \multicolumn{1}{c}{Treated} & \multicolumn{1}{c}{Control} \\
 \hline
 Male  &  \multicolumn{1}{c}{0.60} &  & \multicolumn{1}{c}{0.60}  & \multicolumn{1}{c}{0.60}\\
 Born in 1992  &  \multicolumn{1}{c}{0.05} & & \multicolumn{1}{c}{0.01}  & \multicolumn{1}{c}{0.01}\\
 Born in 1993  &  \multicolumn{1}{c}{0.14} & & \multicolumn{1}{c}{0.12} & \multicolumn{1}{c}{0.12}\\
 Born in 1994  &  \multicolumn{1}{c}{0.22} & & \multicolumn{1}{c}{0.26} & \multicolumn{1}{c}{0.26}\\
 Born in 1995  &  \multicolumn{1}{c}{0.18} & & \multicolumn{1}{c}{0.24} & \multicolumn{1}{c}{0.24}\\
 Born in 1996  &  \multicolumn{1}{c}{0.10} & & \multicolumn{1}{c}{0.17} & \multicolumn{1}{c}{0.17}\\
 Born in 1997  &  \multicolumn{1}{c}{0.07} & & \multicolumn{1}{c}{0.11} & \multicolumn{1}{c}{0.11}\\
 Born in 1998  &  \multicolumn{1}{c}{0.08} &  & \multicolumn{1}{c}{0.09} & \multicolumn{1}{c}{0.09}\\
 Born in January  &  \multicolumn{1}{c}{0.09}  & & \multicolumn{1}{c}{0.10} & \multicolumn{1}{c}{0.10}\\
 Born in February  &  \multicolumn{1}{c}{0.08} & & \multicolumn{1}{c}{0.08} & \multicolumn{1}{c}{0.08}\\
 Born in March  &  \multicolumn{1}{c}{0.08}  & & \multicolumn{1}{c}{0.09} & \multicolumn{1}{c}{0.09}\\
 Born in April  &  \multicolumn{1}{c}{0.08}  & & \multicolumn{1}{c}{0.08} & \multicolumn{1}{c}{0.08}\\
 Born in May  &  \multicolumn{1}{c}{0.09}  & & \multicolumn{1}{c}{0.06} & \multicolumn{1}{c}{0.06}\\
 Born in June  &  \multicolumn{1}{c}{0.09} & & \multicolumn{1}{c}{0.07} & \multicolumn{1}{c}{0.07}\\
 Born in July  &  \multicolumn{1}{c}{0.08}  & & \multicolumn{1}{c}{0.08} & \multicolumn{1}{c}{0.08}\\
 Born in August  &  \multicolumn{1}{c}{0.08}  & & \multicolumn{1}{c}{0.09} & \multicolumn{1}{c}{0.09}\\
 Born in September  &  \multicolumn{1}{c}{0.08}  & & \multicolumn{1}{c}{0.08} & \multicolumn{1}{c}{0.08}\\
 Born in October &  \multicolumn{1}{c}{0.09} & & \multicolumn{1}{c}{0.09} & \multicolumn{1}{c}{0.09}\\
 Born in November  &  \multicolumn{1}{c}{0.09}  & & \multicolumn{1}{c}{0.09} & \multicolumn{1}{c}{0.09}\\
 Born in December &  \multicolumn{1}{c}{0.09}  & & \multicolumn{1}{c}{0.09} & \multicolumn{1}{c}{0.09}\\
 Public School &  \multicolumn{1}{c}{0.68}  & & \multicolumn{1}{c}{0.43} & \multicolumn{1}{c}{0.43}\\
 Subsidized School &  \multicolumn{1}{c}{0.32}  & & \multicolumn{1}{c}{0.57} & \multicolumn{1}{c}{0.57}\\
 3th grade &  \multicolumn{1}{c}{0.07} & & \multicolumn{1}{c}{0.05} & \multicolumn{1}{c}{0.05}\\
 4th grade &  \multicolumn{1}{c}{0.08} & & \multicolumn{1}{c}{0.08} & \multicolumn{1}{c}{0.08}\\
 5th grade &  \multicolumn{1}{c}{0.17} & & \multicolumn{1}{c}{0.17} & \multicolumn{1}{c}{0.17}\\
 6th grade &  \multicolumn{1}{c}{0.31}  & & \multicolumn{1}{c}{0.21} & \multicolumn{1}{c}{0.21}\\
 7th grade &  \multicolumn{1}{c}{0.23}  & & \multicolumn{1}{c}{0.32} & \multicolumn{1}{c}{0.32}\\
 8th grade &  \multicolumn{1}{c}{0.13}  & & \multicolumn{1}{c}{0.17} & \multicolumn{1}{c}{0.17}\\
 Previous grade retention &  \multicolumn{1}{c}{0.39}  & & \multicolumn{1}{c}{0.09} & \multicolumn{1}{c}{0.09}\\
 SIMCE Math &  \multicolumn{1}{c}{-0.53}  & & \multicolumn{1}{c}{-0.26} & \multicolumn{1}{c}{-0.25}\\
 SIMCE Language &  \multicolumn{1}{c}{-0.51}  & & \multicolumn{1}{c}{-0.27} & \multicolumn{1}{c}{-0.26}\\
 Father schooling &  \multicolumn{1}{c}{9.27}  & & \multicolumn{1}{c}{11.37} & \multicolumn{1}{c}{11.42}\\
 Mother schooling &  \multicolumn{1}{c}{9.08}  & & \multicolumn{1}{c}{11.13} & \multicolumn{1}{c}{11.24}\\
 Household income per capita &  \multicolumn{1}{c}{40,720.06}  & & \multicolumn{1}{c}{92,373.58} & \multicolumn{1}{c}{92,297.38}\\
 \hline
\end{tabular}
Notes:  in a neighborhood of the cutoffs, we do exact matching on school, grade attended in 2007, gender, birth month and year, and times of past grade retention, while mean balance matching on SIMCE's score in math and language.  The secondary covariates are father and mother schooling, and household income per capita.
    \end{threeparttable}%
    }
\end{center}
\end{table}

Following Zubizarreta et al. (2018),
%\cite{zubizarreta2018designmatch},
we use cardinality matching to find the largest matched sample that is not only balanced across treatment groups, but is also balanced around the distribution of a target population of interest. Within the window previously determined, Table \ref{table:balance1} compares the means of the treated and control students in the matched sample to the
means of students in the poorest 50\% of households, whereas Table \ref{table:balance2} compares the means of the treated and control students in the representative matched sample to the means of students in the poorest 50\% of households. To find the representative matched sample, we require that the absolute standardized differences in means between the matched treated students and all students in the target population, and between the matched
control students and all students in the target population, are all smaller than 0.05.

Having found a representative matched sample of 283 pairs of students, we assume that treatment is as-if randomly assigned to students conditional on the matched pairs.
Table \ref{table:geneffects} presents the results when generalizing the effect estimates to the population of the poorest 50\% of students with grades in the selected neighborhood, while Tables \ref{table:genpairscrime}, \ref{table:genpairsdrop}, and \ref{table:genpairsret} detail the discordant pairs for committing a crime, dropping out of school, and repeating another grade, respectively.
In the absence of hidden bias, there is no evidence that grade retention causes juvenile crime (the two-sided $p$-value is 0.359) nor dropping out of school ($p$-value = 0.464), whereas there is statistically significant evidence that grade retention reduces the probability of future grade retention ($p$-value = 0.049).

\begin{table}[htbp]
\begin{center}
\caption{Description of the target population and the representative matched sample within the window $[3.5, 4.4]\times [4.3, 4.6]$ for rule 1 and $[3.5, 4.4] \times [4.8, 5.1]$ for rule 2.}
\label{table:balance2}
\scalebox{0.8}{
\begin{threeparttable}[t]
  \centering
 \begin{tabular}{l p{2.2cm} c p{2.2cm} p{2.2cm} }
 \hline
 \multirow{2}{*}{Covariate} &\multicolumn{1}{c}{Full sample mean of the} & & \multicolumn{2}{c}{Matched sample mean} \\
 \cline{4-5}
& \multicolumn{1}{c}{poorest 50\% of households} & &  \multicolumn{1}{c}{Treated} & \multicolumn{1}{c}{Control} \\
 \hline
 Male  &  \multicolumn{1}{c}{0.60} &  & \multicolumn{1}{c}{0.61}  & \multicolumn{1}{c}{0.61}\\
 Born in 1992  &  \multicolumn{1}{c}{0.05} & & \multicolumn{1}{c}{0.04}  & \multicolumn{1}{c}{0.04}\\
 Born in 1993  &  \multicolumn{1}{c}{0.14} & & \multicolumn{1}{c}{0.12} & \multicolumn{1}{c}{0.12}\\
 Born in 1994  &  \multicolumn{1}{c}{0.22} & & \multicolumn{1}{c}{0.20} & \multicolumn{1}{c}{0.20}\\
 Born in 1995  &  \multicolumn{1}{c}{0.18} & & \multicolumn{1}{c}{0.18} & \multicolumn{1}{c}{0.18}\\
 Born in 1996  &  \multicolumn{1}{c}{0.10} & & \multicolumn{1}{c}{0.09} & \multicolumn{1}{c}{0.09}\\
 Born in 1997  &  \multicolumn{1}{c}{0.07} & & \multicolumn{1}{c}{0.08} & \multicolumn{1}{c}{0.08}\\
 Born in 1998  &  \multicolumn{1}{c}{0.08} &  & \multicolumn{1}{c}{0.10} & \multicolumn{1}{c}{0.10}\\
 Born in January  &  \multicolumn{1}{c}{0.09}  & & \multicolumn{1}{c}{0.07} & \multicolumn{1}{c}{0.07}\\
 Born in February  &  \multicolumn{1}{c}{0.08} & & \multicolumn{1}{c}{0.07} & \multicolumn{1}{c}{0.07}\\
 Born in March  &  \multicolumn{1}{c}{0.08}  & & \multicolumn{1}{c}{0.08} & \multicolumn{1}{c}{0.08}\\
 Born in April  &  \multicolumn{1}{c}{0.08}  & & \multicolumn{1}{c}{0.09} & \multicolumn{1}{c}{0.09}\\
 Born in May  &  \multicolumn{1}{c}{0.09}  & & \multicolumn{1}{c}{0.07} & \multicolumn{1}{c}{0.07}\\
 Born in June  &  \multicolumn{1}{c}{0.09} & & \multicolumn{1}{c}{0.07} & \multicolumn{1}{c}{0.07}\\
 Born in July  &  \multicolumn{1}{c}{0.08}  & & \multicolumn{1}{c}{0.07} & \multicolumn{1}{c}{0.07}\\
 Born in August  &  \multicolumn{1}{c}{0.08}  & & \multicolumn{1}{c}{0.08} & \multicolumn{1}{c}{0.08}\\
 Born in September  &  \multicolumn{1}{c}{0.08}  & & \multicolumn{1}{c}{0.10} & \multicolumn{1}{c}{0.10}\\
 Born in October &  \multicolumn{1}{c}{0.09} & & \multicolumn{1}{c}{0.10} & \multicolumn{1}{c}{0.10}\\
 Born in November  &  \multicolumn{1}{c}{0.09}  & & \multicolumn{1}{c}{0.09} & \multicolumn{1}{c}{0.09}\\
 Born in December &  \multicolumn{1}{c}{0.09}  & & \multicolumn{1}{c}{0.09} & \multicolumn{1}{c}{0.09}\\
 Public School &  \multicolumn{1}{c}{0.68}  & & \multicolumn{1}{c}{0.66} & \multicolumn{1}{c}{0.66}\\
 Subsidized School &  \multicolumn{1}{c}{0.32}  & & \multicolumn{1}{c}{0.34} & \multicolumn{1}{c}{0.34}\\
 3th grade &  \multicolumn{1}{c}{0.07} & & \multicolumn{1}{c}{0.07} & \multicolumn{1}{c}{0.07}\\
 4th grade &  \multicolumn{1}{c}{0.08} & & \multicolumn{1}{c}{0.07} & \multicolumn{1}{c}{0.07}\\
 5th grade &  \multicolumn{1}{c}{0.17} & & \multicolumn{1}{c}{0.17} & \multicolumn{1}{c}{0.17}\\
 6th grade &  \multicolumn{1}{c}{0.31}  & & \multicolumn{1}{c}{0.33} & \multicolumn{1}{c}{0.33}\\
 7th grade &  \multicolumn{1}{c}{0.23}  & & \multicolumn{1}{c}{0.25} & \multicolumn{1}{c}{0.25}\\
 8th grade &  \multicolumn{1}{c}{0.13}  & & \multicolumn{1}{c}{0.12} & \multicolumn{1}{c}{0.12}\\
 Previous grade retention &  \multicolumn{1}{c}{0.39}  & & \multicolumn{1}{c}{0.36} & \multicolumn{1}{c}{0.36}\\
 SIMCE Math &  \multicolumn{1}{c}{-0.53}  & & \multicolumn{1}{c}{-0.55} & \multicolumn{1}{c}{-0.51}\\
 SIMCE Language &  \multicolumn{1}{c}{-0.51}  & & \multicolumn{1}{c}{-0.54} & \multicolumn{1}{c}{-0.49}\\
 Father schooling &  \multicolumn{1}{c}{9.27}  & & \multicolumn{1}{c}{9.41} & \multicolumn{1}{c}{9.39}\\
 Mother schooling &  \multicolumn{1}{c}{9.08}  & & \multicolumn{1}{c}{9.20} & \multicolumn{1}{c}{9.20}\\
 Household income per capita &  \multicolumn{1}{c}{40,720.06}  & & \multicolumn{1}{c}{42,641.51} & \multicolumn{1}{c}{43,895.37}\\
 \hline
\end{tabular}

     \item[] Notes: in a neighborhood of the cutoffs, we balance the above covariates around the distribution of a target population of interest.
    \end{threeparttable}%
    }
\end{center}
\end{table}

\begin{table}[h]
\begin{center}
\caption{Generalizing the results in the target population of the poorest 50\% of students with running variables in $[3.5, 4.4]\times [4.3, 4.6]$ for rule 1 and $[3.5, 4.4] \times [4.8, 5.1]$ for rule 2 in the absence of hidden bias.}
\label{table:geneffects}
\scalebox{0.8}{
\begin{threeparttable}[t]
  \centering
 \begin{tabular}{l p{2.2cm}  p{2.2cm}  p{2.2cm} c }
 \hline
 \multirow{2}{*}{Outcome variable} &\multicolumn{2}{c}{Matched sample mean} & \multicolumn{1}{c}{\multirow{2}{*}{$\widehat{\tau}_{\rm{TATE}}$}}  & $H_0: \tau_{\rm{TATE}}=0$ \\
 \cline{2-3}
& \multicolumn{1}{c}{Treated} & \multicolumn{1}{c}{Control} &  & Two-sided $p$-value \\
 \hline
 Committing a crime  &  \multicolumn{1}{c}{0.078} & \multicolumn{1}{c}{0.095} & \multicolumn{1}{c}{-0.017}  & 0.359\\
 Future grade retention  &  \multicolumn{1}{c}{0.477} & \multicolumn{1}{c}{0.555} & \multicolumn{1}{c}{-0.078}  & 0.049\\
 Dropping out of school  &  \multicolumn{1}{c}{0.191} & \multicolumn{1}{c}{0.173} & \multicolumn{1}{c}{0.018} & 0.464\\
 \hline
\end{tabular}
   Notes: we report the two-sided $p$-value for McNemar's test of no effect on the TATE.
    \end{threeparttable}%
    }
\end{center}
\end{table}

\begin{table}[h]
\begin{center}
\caption{Generalizing the results: juvenile crime indicators in the representative matched pairs sample. The table counts pairs, not students.}
\label{table:genpairscrime}
\scalebox{0.8}{
  \centering
 \begin{tabular}{l c  c}
 \hline
  &\multicolumn{2}{c}{Passed}  \\
 \cline{2-3}
Repeated & Juvenile crime = 0 & Juvenile crime = 1 \\
 \hline
 Juvenile crime = 0  &  238 & 23\\
 Juvenile crime = 1  &  18 & 4 \\
 \hline
\end{tabular}
}
\end{center}
\end{table}

\begin{table}[h]
\begin{center}
\caption{Generalizing the results: dropping out of school indicators in the representative matched pairs sample. The table counts pairs, not students.}
\label{table:genpairsdrop}
\scalebox{0.8}{
  \centering
 \begin{tabular}{l c  c}
 \hline
  &\multicolumn{2}{c}{Passed}  \\
 \cline{2-3}
Repeated & Dropped out = 0 & Dropped out = 1 \\
 \hline
 Dropped out = 0  &  198 & 31\\
 Dropped out = 1  &  36 & 18 \\
 \hline
\end{tabular}
}
\end{center}
\end{table}

\begin{table}[h]
\begin{center}
\caption{Generalizing the results: future grade retention indicators in the representative matched pairs sample. The table counts pairs, not students.}
\label{table:genpairsret}
\scalebox{0.8}{
  \centering
 \begin{tabular}{l c  c}
 \hline
  &\multicolumn{2}{c}{Passed}  \\
 \cline{2-3}
Repeated & Future retention = 0 & Future retention = 1 \\
 \hline
 Future retention = 0  &  68 & 80\\
 Future retention = 1  &  58 & 77 \\
 \hline
\end{tabular}
}
\end{center}
\end{table}

%%%%%%%%%%%%%%%%%%%%%%%%%%%%%%%%%%%%%%%%%%%
%%%%%%%%%%%%%%%%%%%%%%%%%%%%%%%%%%%%%%%%%%%
%%%%%%%%%%%%%%%%%%%%%%%%%%%%%%%%%%%%%%%%%%%
\clearpage
\pagebreak
\section*{Appendix G: Selecting a neighborhood}

%%%%%%%%%%%%%%%%%%%%%%%%%%%%%%%%%%%%%%%%%%%
%%%%%%%%%%%%%%%%%%%%%%%%%%%%%%%%%%%%%%%%%%%
\subsection*{Implementation of the design-based approach}

In the design-based approach, we can let $X_i^{\textrm{test}} = Y_i^{\textrm{lagged}}$, the lagged outcome (i.e., the outcome variable measured before treatment exposure), but $X_i^{\textrm{test}}$ might as well be a lagged running variable or a secondary covariate.
We see $Y_i^{\textrm{lagged}}$ as a proxy of the potential outcome under control (see Chapter 21 of \citealp{imbens2015causal}, for a discussion).
If Assumption 1\textcolor{black}{a} holds, then $Y_i^{\textrm{lagged}}$ should be balanced across treatment groups for all units with $N_i=1$, after adjusting for $\boldsymbol{X}_i$.
We thus assess the  plausibility of Assumption 1\textcolor{black}{a} by checking whether the following mean independence condition holds
\begin{eqnarray}\label{laggoutcome}
  \mathbb{E}\bigl\{g(Y_i^{\textrm{lagged}})\ \big| \ \boldsymbol{R}_{i1},...,\boldsymbol{R}_{i\mathrm{J}}, \ \boldsymbol{X}_i, \ N_i=1\bigl\}=\mathbb{E}\bigl\{g(Y_i^{\textrm{lagged}})\ \big| \ \boldsymbol{X}_i, \ N_i=1\bigl\},
\end{eqnarray}
for any function $g(\cdot)$ with finite moments.
Here, $\boldsymbol{X}_i$ does not include $Y_i^{\textrm{lagged}}$.

There are several ways of testing (\ref{laggoutcome}), both in parametric and nonparametric setups.
In a parametric setup, we can regress $g(Y_i^{\textrm{lagged}})$ on the running variables and $\boldsymbol{X}_i$ for the units with $N_i=1$ and test the null hypothesis that all the coefficients of the running variables are jointly zero.
We select the largest neighborhood where we fail to reject the null.
However, although within this parametric setup we can posit a flexible model for the conditional mean function, mean independence is stronger than regression independence (for a given family of regression functions).
Thus, in order to directly test the mean independence condition in (\ref{laggoutcome}), we can instead use nonparametric regression.
For a comprehensive review and methods for related nonparametric regression tests, see \cite{fan1996consistent}, \cite{delgado2001significance}, and \cite{lavergne2015significance}.
Essentially, for units with $N_i=1$, in this nonparametric setup we test
$$H_0: \ \ \mathbb{E}\big\{g(Y_i^{\textrm{lagged}}) \ \big| \ \boldsymbol{R}_{i1},...,\boldsymbol{R}_{i\mathrm{J}}, \ \boldsymbol{X}_i\big\}=\mathbb{E}\big\{g(Y_i^{\textrm{lagged}}) \ \big| \ \boldsymbol{X}_i\big\} \ \ \mathrm{a.s.}$$
versus the corresponding alternative hypothesis
$$H_1: \ \ \Pr \big[\mathbb{E}\big\{g(Y_i^{\textrm{lagged}})-\mathbb{E}(g(Y_i^{\textrm{lagged}}) \ \big| \ \boldsymbol{X}_i) \ \big| \ \boldsymbol{R}_{i1},...,\boldsymbol{R}_{i\mathrm{J}}, \ \boldsymbol{X}_i\big\}=0\big] \ \ \mathrm{a.s.}$$
Despite the obvious appeal of this nonparametric testing procedure, its main disadvantage is its limited power in practice.
To overcome this limitation, alternatively we can use the matching approach to assess (\ref{laggoutcome}) by checking
%\begin{align}\label{balance2}
%\mathbb{E}\bigl[\mathbb{E}\bigl\{g(Y_i^{\textrm{lagged}})\ \big| \ Z_i=1, \ \boldsymbol{X}_i, \  N_i=1\bigl\}
%& -\mathbb{E}\bigl\{g(Y_i^{\textrm{lagged}})\ \big| Z_i=0, \ \boldsymbol{X}_i, \ N_i=1\bigl\}\big| N_i=1\bigl]=0.
%\end{align}
\vspace{-.25cm}
\begin{widerequation}\label{balance2}
\mathbb{E}\bigl[\mathbb{E}\bigl\{g(Y_i^{\textrm{lagged}})\ \big| \ Z_i=1, \ \boldsymbol{X}_i, \  N_i=1\bigl\} -\mathbb{E}\bigl\{g(Y_i^{\textrm{lagged}})\ \big| Z_i=0, \ \boldsymbol{X}_i, \ N_i=1\bigl\}\big| N_i=1\bigl]=0. 
\end{widerequation}

Here, the idea is to use matching to adjust for $\boldsymbol{X}_i$ and to conduct a sequence of balance tests for different neighborhoods, selecting the largest one that is compatible with (\ref{balance2}).

%Assessing balance of pretreatment variables between treatment groups to evaluate the plausibility of a critical assumption is a common strategy in observational studies.
%Typically, this strategy has been used to select a matched sample when analyzing observational data under strong ignorability (\citealt{imbens2015causal}, Chapter 21).
Assessing balance of pretreatment variables between treatment groups to evaluate the plausibility of the unconfoundedness assumption is a common strategy in observational studies (\citealp{imbens2015causal}, Chapter 21).
%To our knowledge, the idea of selecting the neighborhood based on covariates in a discontinuity design was first proposed by \cite{cattaneo2015randomization}.
In the local randomization framework, \cite{cattaneo2015randomization} propose using a sequence of tests for balance on covariates other than the running variable to assess the assumption of local randomization and select the neighborhood (see also \citealp{cattaneo2017comparing}).
In a Bayesian framework, \cite{li2015evaluating} introduce a sequence of tests for covariate balance to choose a neighborhood where the local randomization assumption is plausible, while \cite{mattei2016regression} employ randomization-based tests that adjust for multiple comparisons.
We adapt these approaches to a setting where the researcher counts with two set of covariates besides the running variables: a first set involved in the conditional statement of Assumption 1\textcolor{black}{a}, and a second set that allows us to select the largest neighborhood for analysis by sequentially assessing the balance condition presented in Equation (15) in Section 5.
%In this paper we propose a procedure that employs two sets of covariates other than the running variables to simultaneously adjust and test for covariate balance in order to select the neighborhood for analysis in a discontinuity design under local strong ignorability.
These approaches are implemented in the design stage of the study, without using the outcomes and therefore blinding the investigator from the study results until after the selection of the neighborhood.

%%%%%%%%%%%%%%%%%%%%%%%%%%%%%%%%%%%%%%%%%%%
%%%%%%%%%%%%%%%%%%%%%%%%%%%%%%%%%%%%%%%%%%%
\subsection*{A semi-design-based approach}
%\label{sec_semidesign}
%
%The previous approach does not use the outcomes and in that sense is design-based (Imbens and Rubin 2015, Chapter 21).
The neighborhood selection approach presented in Section 5 does not use the outcomes and in that sense is design-based (Imbens and Rubin 2015, Chapter 21).
The approach that follows, by contrast, uses the outcomes, but in a split, smaller planning sample (different from the analysis sample) and in that sense is semi-design-based (\citealt{rosenbaum2010design1}, Chapter 18).
This approach is in the spirit of \cite{heller2009split} who use a planning sample to guide the design of an observational study.
For outcome analyses and estimation of treatment effects, they discard this planning sample to preclude the use of the same outcome data twice (in the design stage and also in the analysis stage of the study).
We adapt this idea to our framework to assess the plausibility of the assumptions and select the neighborhood.
This approach selects the neighborhood from the planning sample using the outcomes and then uses this neighborhood in the analysis sample for estimation of the average treatment effects.

Consider splitting the available study sample at random into two parts, one planning sample of size $n_p$ and an analysis sample of size $n_a$ with $n_p+n_a=n$.
Specifically, given a neighborhood, Assumption 1a implies the next two conditions that can be evaluated using the planning sample
\begin{equation}\label{sd1}
  \mathbb{E}\bigl[g(Y_i(1))\ \big| \ \boldsymbol{R}_{i1},...,\boldsymbol{R}_{i\mathrm{J}}, \ \boldsymbol{X}_i, \ Z_i=1,\ N_i=1\bigl]=\mathbb{E}\bigl[g(Y_i(1))\ \big| \ \boldsymbol{X}_i, \ Z_i=1,\ N_i=1 \bigl],
\end{equation}
\begin{equation}\label{sd2}
\mathbb{E}\bigl[g(Y_i(0))\ \big| \ \boldsymbol{R}_{i1},...,\boldsymbol{R}_{i\mathrm{J}}, \ \boldsymbol{X}_i, \ Z_i=0,\ N_i=1\bigl]=\mathbb{E}\bigl[g(Y_i(0))\ \big| \ \boldsymbol{X}_i, \ Z_i=0,\ N_i=1\bigl],
\end{equation}
where $g(\cdot)$ is an arbitrary function of the covariates with finite moments. In order to test whether the conditions $(\ref{sd1})$ and $(\ref{sd2})$ hold, one can employ any of the three empirical strategies we have mentioned in the previous subsection, namely, a parametric model, a nonparametric significance testing procedure, or a matching-style approach.

Although splitting the sample can reduce the power in the analysis stage, we think that it is an objective way of assessing the critical assumptions required to conduct credible causal inference.
Moreover, if the sample size is large enough, the consideration of a smaller planning sample should not be an issue.
Finally, as \cite{heller2009split} point out, it is recommended to consider repeated splitting samples in order to check the stability of the results.

%%%%%%%%%%%%%%%%%%%%%%%%%%%%%%%%%%%%%%%%%%%
%%%%%%%%%%%%%%%%%%%%%%%%%%%%%%%%%%%%%%%%%%%
\subsection*{Other practical considerations}

A natural question to ask is how to expand the neighborhood.
In typical discontinuity designs with only one running variable, the neighborhood is expanded symmetrically in both directions of the cutoff, but this is a simplification and the neighborhood may be expanded asymmetrically in one direction first.
In more complex designs with many running variables and multiple treatment rules, like in our running example, there are more directions in which the neighborhood can be expanded first (in our running example, there are eight such directions).
In our running example, it is more plausible to believe that, after controlling for the observed covariates, students will be comparable in a larger neighborhood of the cutoff of a single school subject ($R_{11}$ or $R_{21}$) than in a neighborhood of equal size of the average across all subjects ($R_{12}$ and $R_{22}$).
For this reason, we first expand the neighborhoods of $R_{11}$ and $R_{21}$.
For $R_{11}$ and $R_{21}$ we expand the neighborhood symmetrically for simplicity, but this does not need to be the case in general.

%%%%%%%%%%%%%%%%%%%%%%%%%%%%%%%%%%%%%%%%%%%
%%%%%%%%%%%%%%%%%%%%%%%%%%%%%%%%%%%%%%%%%%%
%%%%%%%%%%%%%%%%%%%%%%%%%%%%%%%%%%%%%%%%%%%
\clearpage
\pagebreak
\section*{Appendix H: Varying the size of the neighborhood}

In this appendix, we analyze the stability of these results to varying the neighborhood size. The results are presented in two subsections. In the first subsection, we modify $\underline{\delta}$ and $\overline{\delta}$ to increase the size of the neighborhood for the lowest and second lowest grades in rules 1 and 2, respectively. In the second subsection, we increase the size of the neighborhood for the average grade. As can be seen in the tables, our findings are stable to varying the size of the neighborhood.

\subsection*{Modifying the neighborhood for the lowest and second lowest grades}

\begin{table}[htbp]
\begin{center}
\caption{Neighborhood: Rule 1 $\text{[3.6, 4.3]} \times \text{[4.3, 4.6]}$ and Rule 2 $\text{[3.6, 4.3]} \times \text{[4.8, 5.1]}$}
\label{table:table3}
\scalebox{0.8}{
\begin{threeparttable}[t]
\centering
\begin{tabular}{l p{2.2cm}  p{2.2cm}  p{2.2cm} c }
\hline
\multirow{2}{*}{Outcome variable} &\multicolumn{2}{c}{Matched sample mean} & \multicolumn{1}{c}{\multirow{2}{*}{$\widehat{\tau}_{\rm{NATE}}$}}  & $H_0: \tau_{\rm{NATE}}=0$ \\
\cline{2-3}
& \multicolumn{1}{c}{Treated} & \multicolumn{1}{c}{Control} &  & One-sided $p$-value \\
\hline
Committing a crime  &  \multicolumn{1}{c}{0.058} & \multicolumn{1}{c}{0.058} & \multicolumn{1}{c}{0.000}  & \phantom{<}0.459\\
Repeating another grade  &  \multicolumn{1}{c}{0.452} & \multicolumn{1}{c}{0.565} & \multicolumn{1}{c}{-0.113}  & <0.001\\
 Dropping out of school  &  \multicolumn{1}{c}{0.093} & \multicolumn{1}{c}{0.092} & \multicolumn{1}{c}{\phantom{-}0.001} & \phantom{<}0.430\\
\hline
\end{tabular}
\end{threeparttable}
}
\end{center}
\end{table}

\begin{table}[htbp]
\begin{center}
\caption{Neighborhood: Rule 1 $\text{[3.4, 4.5]} \times \text{[4.3, 4.6]}$ and Rule 2 $\text{[3.4, 4.5]} \times \text{[4.8, 5.1]}$}
\label{table:table3}
\scalebox{0.8}{
\begin{threeparttable}[t]
\centering
\begin{tabular}{l p{2.2cm}  p{2.2cm}  p{2.2cm} c }
\hline
\multirow{2}{*}{Outcome variable} &\multicolumn{2}{c}{Matched sample mean} & \multicolumn{1}{c}{\multirow{2}{*}{$\widehat{\tau}_{\rm{NATE}}$}}  & $H_0: \tau_{\rm{NATE}}=0$ \\
\cline{2-3}
& \multicolumn{1}{c}{Treated} & \multicolumn{1}{c}{Control} &  & One-sided $p$-value \\
\hline
Committing a crime  &  \multicolumn{1}{c}{0.059} & \multicolumn{1}{c}{0.059} & \multicolumn{1}{c}{0.000}  & \phantom{<}0.527\\
Repeating another grade  &  \multicolumn{1}{c}{0.467} & \multicolumn{1}{c}{0.557} & \multicolumn{1}{c}{-0.091}  & <0.001\\
 Dropping out of school  &  \multicolumn{1}{c}{0.100} & \multicolumn{1}{c}{0.102} & \multicolumn{1}{c}{-0.001} & \phantom{<}0.419\\
\hline
\end{tabular}
\end{threeparttable}
}
\end{center}
\end{table}

\begin{table}[htbp]
\begin{center}
\caption{Neighborhood: Rule 1 $\text{[3.3, 4.6]} \times \text{[4.3, 4.6]}$ and Rule 2 $\text{[3.3, 4.6]} \times \text{[4.8, 5.1]}$}
\label{table:table3}
\scalebox{0.8}{
\begin{threeparttable}[t]
\centering
\begin{tabular}{l p{2.2cm}  p{2.2cm}  p{2.2cm} c }
\hline
\multirow{2}{*}{Outcome variable} &\multicolumn{2}{c}{Matched sample mean} & \multicolumn{1}{c}{\multirow{2}{*}{$\widehat{\tau}_{\rm{NATE}}$}}  & $H_0: \tau_{\rm{NATE}}=0$ \\
\cline{2-3}
& \multicolumn{1}{c}{Treated} & \multicolumn{1}{c}{Control} &  & One-sided $p$-value \\
\hline
Committing a crime  &  \multicolumn{1}{c}{0.060} & \multicolumn{1}{c}{0.061} & \multicolumn{1}{c}{-0.001}  & \phantom{<}0.410\\
Repeating another grade  &  \multicolumn{1}{c}{0.474} & \multicolumn{1}{c}{0.570} & \multicolumn{1}{c}{-0.096}  & <0.001\\
 Dropping out of school  &  \multicolumn{1}{c}{0.104} & \multicolumn{1}{c}{0.109} & \multicolumn{1}{c}{-0.005} & \phantom{<}0.288\\
\hline
\end{tabular}
\end{threeparttable}
}
\end{center}
\end{table}

\subsection*{Modifying the neighborhood for the average grade}

\begin{table}[htbp]
\begin{center}
\caption{Neighborhood: Rule 1 $\text{[3.5, 4.4]} \times \text{[4.4, 4.5]}$ and Rule 2 $\text{[3.5, 4.4]} \times \text{[4.9, 5.0]}$}
\label{table:table3}
\scalebox{0.8}{
\begin{threeparttable}[t]
\centering
\begin{tabular}{l p{2.2cm}  p{2.2cm}  p{2.2cm} c }
\hline
\multirow{2}{*}{Outcome variable} &\multicolumn{2}{c}{Matched sample mean} & \multicolumn{1}{c}{\multirow{2}{*}{$\widehat{\tau}_{\rm{NATE}}$}}  & $H_0: \tau_{\rm{NATE}}=0$ \\
\cline{2-3}
& \multicolumn{1}{c}{Treated} & \multicolumn{1}{c}{Control} &  & One-sided $p$-value \\
\hline
Committing a crime  &  \multicolumn{1}{c}{0.056} & \multicolumn{1}{c}{0.043} & \multicolumn{1}{c}{0.013}  & \phantom{<}0.191\\
Repeating another grade  &  \multicolumn{1}{c}{0.365} & \multicolumn{1}{c}{0.586} & \multicolumn{1}{c}{-0.221}  & <0.001\\
 Dropping out of school  &  \multicolumn{1}{c}{0.082} & \multicolumn{1}{c}{0.117} & \multicolumn{1}{c}{-0.034} & \phantom{<}0.100\\
\hline
\end{tabular}
\end{threeparttable}
}
\end{center}
\end{table}

\begin{table}[htbp]
\begin{center}
\caption{Neighborhood: Rule 1 $\text{[3.5, 4.4]} \times \text{[4.2, 4.7]}$ and Rule 2 $\text{[3.5, 4.4]} \times \text{[4.7, 5.2]}$}
\label{table:table3}
\scalebox{0.8}{
\begin{threeparttable}[t]
\centering
\begin{tabular}{l p{2.2cm}  p{2.2cm}  p{2.2cm} c }
\hline
\multirow{2}{*}{Outcome variable} &\multicolumn{2}{c}{Matched sample mean} & \multicolumn{1}{c}{\multirow{2}{*}{$\widehat{\tau}_{\rm{NATE}}$}}  & $H_0: \tau_{\rm{NATE}}=0$ \\
\cline{2-3}
& \multicolumn{1}{c}{Treated} & \multicolumn{1}{c}{Control} &  & One-sided $p$-value \\
\hline
Committing a crime  &  \multicolumn{1}{c}{0.055} & \multicolumn{1}{c}{0.051} & \multicolumn{1}{c}{0.004}  & \phantom{<}0.206\\
Repeating another grade  &  \multicolumn{1}{c}{0.458} & \multicolumn{1}{c}{0.551} & \multicolumn{1}{c}{-0.092}  & <0.001\\
 Dropping out of school  &  \multicolumn{1}{c}{0.097} & \multicolumn{1}{c}{0.083} & \multicolumn{1}{c}{\phantom{-}0.013} & \phantom{<}0.040\\
\hline
\end{tabular}
\end{threeparttable}
}
\end{center}
\end{table}

\begin{table}[htbp]
\begin{center}
\caption{Neighborhood: Rule 1 $\text{[3.4, 4.5]} \times \text{[4.2, 4.7]}$ and Rule 2 $\text{[3.4, 4.5]} \times \text{[4.7, 5.2]}$}
\label{table:table3}
\scalebox{0.8}{
\begin{threeparttable}[t]
\centering
\begin{tabular}{l p{2.2cm}  p{2.2cm}  p{2.2cm} c }
\hline
\multirow{2}{*}{Outcome variable} &\multicolumn{2}{c}{Matched sample mean} & \multicolumn{1}{c}{\multirow{2}{*}{$\widehat{\tau}_{\rm{NATE}}$}}  & $H_0: \tau_{\rm{NATE}}=0$ \\
\cline{2-3}
& \multicolumn{1}{c}{Treated} & \multicolumn{1}{c}{Control} &  & One-sided $p$-value \\
\hline
Committing a crime  &  \multicolumn{1}{c}{0.056} & \multicolumn{1}{c}{0.055} & \multicolumn{1}{c}{0.001}  & \phantom{<}0.379\\
Repeating another grade  &  \multicolumn{1}{c}{0.474} & \multicolumn{1}{c}{0.545} & \multicolumn{1}{c}{-0.071}  & <0.001\\
 Dropping out of school  &  \multicolumn{1}{c}{0.096} & \multicolumn{1}{c}{0.085} & \multicolumn{1}{c}{\phantom{-}0.045} & \phantom{<}0.040\\
\hline
\end{tabular}
\end{threeparttable}
}
\end{center}
\end{table}

\begin{table}[htbp]
\begin{center}
\caption{Neighborhood: Rule 1 $\text{[3.3, 4.6]} \times \text{[4.1, 4.8]}$ and Rule 2 $\text{[3.3, 4.6]} \times \text{[4.6, 5.3]}$}
\label{table:table3}
\scalebox{0.8}{
\begin{threeparttable}[t]
\centering
\begin{tabular}{l p{2.2cm}  p{2.2cm}  p{2.2cm} c }
\hline
\multirow{2}{*}{Outcome variable} &\multicolumn{2}{c}{Matched sample mean} & \multicolumn{1}{c}{\multirow{2}{*}{$\widehat{\tau}_{\rm{NATE}}$}}  & $H_0: \tau_{\rm{NATE}}=0$ \\
\cline{2-3}
& \multicolumn{1}{c}{Treated} & \multicolumn{1}{c}{Control} &  & One-sided $p$-value \\
\hline
Committing a crime  &  \multicolumn{1}{c}{0.059} & \multicolumn{1}{c}{0.047} & \multicolumn{1}{c}{0.011}  & \phantom{<}0.004\\
Repeating another grade  &  \multicolumn{1}{c}{0.481} & \multicolumn{1}{c}{0.513} & \multicolumn{1}{c}{-0.031}  & \phantom{-}0.001\\
 Dropping out of school  &  \multicolumn{1}{c}{0.099} & \multicolumn{1}{c}{0.079} & \multicolumn{1}{c}{\phantom{-}0.020} & \phantom{<}0.000\\
\hline
\end{tabular}
\end{threeparttable}
}
\end{center}
\end{table}

%%%%%%%%%%%%%%%%%%%%%%%%%%%%%%%%%%%%%%%%%%%
%%%%%%%%%%%%%%%%%%%%%%%%%%%%%%%%%%%%%%%%%%%
%%%%%%%%%%%%%%%%%%%%%%%%%%%%%%%%%%%%%%%%%%%
\clearpage
\pagebreak
\section*{Appendix I: Assessing balance under the continuity-based framework}

Under the continuity-based framework, we also find significant imbalances at the cutoff.
Here, we analyze each of the two treatment rules separately, and within each rule, collapse the two running variables into a single
dimension by restricting the sample to students below one of the cutoffs.
For example, for Rule 1, we restrict the sample to students with a grade below 4 in one subject and estimated the average effect of treatment at average grade across all subjects of 4.5 using the methods proposed by \cite{calonico2014robust}.
The results are presented in Table \ref{tab1}, which shows significant differences in the math and language test scores.
%This rules out the validity of the assumptions of the traditional continuity and local randomization frameworks.

%Regarding (a), we explored your suggestion and implemented designs (1) and (3).
%Under both the continuity- and local randomization-based frameworks, 
%We found that in the smallest possible neighborhoods around the cutoffs, the basic assumptions of the traditional continuity and local randomization frameworks do not hold because key covariates are not balanced.
%For this, \textcolor{blue}{we implemented the methods for estimation and inference in} \cite{calonico2014robust}.

\begin{table}[htbp]
\begin{center}
\caption{}
\label{tab1}
\scalebox{0.8}{
\begin{tabular}{lcccc}
\hline
 & \multicolumn{2}{c}{Sample of students with $R_{11}<4$} & \multicolumn{2}{c}{Sample of students with $R_{21}<4$}\\
 & \multicolumn{2}{c}{Effect at $R_{12}=4.5$} & \multicolumn{2}{c}{Effect at $R_{22}=5$}\\
 
              & Math  &   Language  & Math & Language\\
              \hline
RD point estimator & -0.060 & -0.057 & -0.114 & -0.103\\
Standard error & (0.015) & (.0139) & (0.035) & (0.035) \\
Robust $p$-value & <0.001 & <0.001 & 0.003 & 0.008 \\
95\% C.I. & [-0.010, -0.033] & [-0.093, -0.032] & [-0.204, -0.043] & [-0.192, -0.029] \\
\hline
\end{tabular}
}
\end{center}
\end{table}

%%%%%%%%%%%%%%%%%%%%%%%%%%%%%%%%%%%%%%%%%%%
%%%%%%%%%%%%%%%%%%%%%%%%%%%%%%%%%%%%%%%%%%%
%%%%%%%%%%%%%%%%%%%%%%%%%%%%%%%%%%%%%%%%%%%
%\pagebreak
%\onehalfspacing
%\bibliography{mybibliography18}
%\bibliographystyle{asa}